\documentclass[journal,singlecolumn]{IEEEtran}
\usepackage{graphicx}
\usepackage{amsmath}
\usepackage{amsfonts}
\usepackage{paralist}
\usepackage{arabtex}
\usepackage{comment}
\usepackage[utf8]{inputenc}
\usepackage{multirow}

\usepackage{textcomp,marvosym}

\usepackage{nameref,hyperref}

\begin{document}
\title{Towards a computational model of social norms}
\author{Ladislau~B{\"o}l{\"o}ni,~\IEEEmembership{Senior Member,~IEEE},
   Taranjeet~Singh~Bhatia,~\IEEEmembership{Student Member,~IEEE,}
   Saad~Ahmad~Khan,~\IEEEmembership{Student Member,~IEEE,}
   Jonathan Streater and~Stephen M. Fiore
        \thanks{S. A. Khan is with the Dept. of ECE, University of Central Florida,
        FL, 32826 USA e-mail: }
        \thanks{T. S. Bhatia is with the Dept. of CS, University of Central Florida,
        FL, 32826 USA e-mail: (see http://www.eecs.ucf.edu/~tsbhaita).}
        \thanks{J. Streater was with the Institute for Simulation and Training,
        University of Central Florida}
        \thanks{S. M. Fiore is with the Institute for Simulation and Training,
        University of Central Florida}
        \thanks{L. B{\"o}l{\"o}ni is with the Dept. of CS, University of Central Florida,
        FL, 32826 USA e-mail: (see http://www.eecs.ucf.edu/~lboloni).}}

\maketitle

\begin{abstract}

We describe a computational model of social norms based on identifying values that a certain culture finds desirable such as dignity, generosity and politeness. The model quantifies these values in the form of Culture-Sanctioned Social Metrics (CSSMs) and treats social norms as the requirement to maximize these metrics from the perspective of the self, peers and public. This model can be used to create realistic social simulations, to explain or predict human behavior in specific scenarios, or as a component of robots or agents that need to interact with humans in specific social-cultural settings. We validate the model by using it to represent a complex deception scenario and showing that it can yield non-trivial insights such as the explanation of apparently irrational human behavior.

\end{abstract}

%
%

\newtheorem{definition}{Definition}
\newtheorem{conjecture}{Conjecture}

\section{Introduction}

There are many applications where we need to computationally reason about social norms. Agent-based social simulations that do not account for the norms of the modeled culture will generate misleading results. Games and interactive narratives that do not model social norms will have actors behaving in an unrealistic, brusque way, preventing the ``suspension of disbelief'' and hindering the educational mission of ``serious games''. In human-agent or human-robot interaction, an artificial entity that does not follow the appropriate social norms can annoy or confuse the human participant, lowering the quality of the experience or even preventing the successful conclusion of the interaction.

The traditional way to account for social norms in such systems is to employ a knowledgeable scenario designer to carefully script the interaction such that the behaviors account for social norms. This is a labor intensive process that limits the number of possible traversals of a given scenario and does not scale well, as the scripts need to be developed from scratch for every new scenario.

In this paper, we introduce a general purpose computational framework for reasoning about social norms. We aim the model to be general enough that its basic computational engine to be reusable across arbitrary scenarios. The model should take into account the fact that social norms are often expressed in terms of the perceptions of the interaction partners, peers or public. Furthermore, the model needs to account for different cultures, as well as situations of cross-cultural interactions.

We start from the observation that different cultures {\em sanction} certain {\em values} as desirable. Our model assumes that these values can be assigned numerical metrics. Actions taken in social settings positively or negatively impact these metrics. The behavior of actors in social scenarios will take into account the objective to maximize the relevant social metrics - but naturally, upholding social norms is not the only goal of the actors.

The reminder of this paper is organized as follows. We start with a survey of  approaches for modeling social norms from social sciences, and justify the need for our model by pointing out that the objective of these models are different from ours - they aim for a qualitative description of behavior rather than a generative computational model. We also survey work done in computational sciences in generating social behaviors.

In the next section of the paper we introduce the components of our model. First, we formalize the components of a social scenario. Second we introduce Culture Sanctioned Social Metrics (CSSMs), the numerical measures that describe cultural values. As the evaluation of values by different social actors depend on the their interpretation of the actions by other actors, we also introduce a model for the Concrete Beliefs (CBs) of the actors relevant to the social scenario. The next section is dedicated to techniques through which we can represent the changes in CSSMs and CBs.

The final step in proposing a computation model is its validation. How can we validate a computational model of social norms? While formal rigor and ease of computational implementation can be easily verified by construction, validating a computational model applied to human behavior is difficult and not subject to analytical proofs. We argue that the usefulness of computational model can be validated if we prove its explanatory and predictive power in non-trivial scenarios. We use the proposed approach to model a real world social scenario, the Spanish Steps flower selling scam, We show that our model allows us to explain and predict the apparently irrational behavior of the participants.

%
%

\section{Related work}
\label{sec:RelatedWork}

Many theories developed in the psychological and social sciences account for social norms in the study of human behavior as it exists in current or past societies. Some of these theories aim for qualitative descriptions that facilitate human understanding, but are difficult to model computationally. Other theories use quantitative, numerical metrics to characterize human behavior and are thus closer to our objective of {\em generating} behavior in the presence of social norms. The quantified concepts differ from model to model. In the following we will briefly survey models of social norms that emerge from quantitative theories of emotions and quantitative models of culture.

\subsection*{Models of social norms in theories of emotions}

Culture and social norms can affect the emotions of humans in several ways. In some cases, social norms directly prescribe emotional reactions, by specifying the {\em ideal affect} a person should try to achieve~\cite{Tsai-2007-IdealAffect}. More directly, however, all cultures expect people to have an emotional response towards the fulfillment or violation of the social norms.

The {\em appraisal theory of emotions}, one of the most influential emotion models in psychology, derives emotions from the meaning that an individual attaches to an event. This meaning can often be equated with its adaptive or coping value. There are several alternative formulations of appraisal theory, giving slightly different definitions for emotions associated with social norms. For instance, the Orthony, Clore and Collins theory of emotion triggering (OCC)~\cite{Ortony-1990-CognitiveStructureOfEmotions} introduces the separate class of {\em attribution emotions} that are attached to events that uphold or violate social norms. On the other hand, the appraisal theory of Roseman~\cite{Roseman-2001-ModelOfAppraisal} integrates these into a class called {\em self-caused emotions}, which can be positive (pride) or negative (regret, guilt or shame) function of their instrumentality and control potential.

Another possible perspective to emotions is the economic one: Jon Elster associates social norms~\cite{Elster-1998-EmotionsEconomic} with rules of behavior that are {\em not} outcome-oriented.

There is a significant body of work on the computational models of emotions (surveyed, for instance in \cite{Hudlicka-2011-CompModelsOfEmotions} and \cite{Marsella-2010-ComputationalModelsOfEmotion}). Some researchers were motivated by the desire to implement emotional agents or robots, while others had the explicit goal of building a formal qualitative or quantitative theory. Many, though not all, of these models implement or are inspired by the OCC model.

Eliott's Affective Reasoner~\cite{Elliott-1992-AffectiveReasoner} used a frame-based representation loosely anchored in the OCC theory. It performed reasoning about emotion generation, action generation from emotional states as well as backward reasoning from observations to the likely emotions that generated them. The EMA architecture of Gratch and Marsella~\cite{GratchMarsella-2004-COGSYS} improved this model, among others with representations for the coping ability of humans. In EMA, norms are encoded only indirectly as preferences over action outcomes, effectively integrating them into means-ends reasoning (with the possibility of representing less desirable means with negative side effects). The Oz project~\cite{Bates-1992-Integrating,Loyall-1997-Thesis,Reilly-1996-Thesis} introduced animated characters that included (among other components of the OCC model) also simple attribution emotions with respect to moral beliefs and personal standards of performance. Examples include ``do-not-cause-my-goals-to-fail'' and ``help-my-goals-succeed''.
The TABASCO agent architecture~\cite{StallerPetta-1998-TABASCO} put the emotional appraisal at the center of the agent, with perception and appraisal happening through the hierarchy of conceptual, schematic and sensory processing, in its turn generating actions at the conceptual, schematic and motor levels. El Nasr et al.~\cite{ElNasr-2000-FLAME} developed a fuzzy logic based model of emotions, which scores emotion intensities based on formulas from~\cite{Price-1985-QuantitativeExperiential}. The authors also associate a reinforcement learning component to the model. For learning attributive emotions the authors assume that the agent already has a numerical metric that quantifies the way the norms are affected by actions, but do not discuss in detail where this number is acquired from. The SCREAM system~\cite{Prendinger-2002-SCREAM} was a scripting tool for creating animated webpage characters with emotional reactions. An interesting aspect of this model is that it combines inspiration from the OCC theory with the representation of social power and social distance from the Brown and Levinson politeness theory~\cite{Brown-1987-Politeness}. Adam et al.~\cite{Adam-2009-LogicalFormalizationOCC} used modal logic to formalize large parts of the OCC theory. This includes a representation of the attribution emotions of pride / shame and admiration / reproach. These models include two intensity variables: one for the {\em strength of unit} which describes how much the agent identifies itself with the author of the action and the {\em expectation deviation} which is related to the difference of the action from what is normally expected from the agent. The approach, however, does not go into detail about the modeling of the norm, simply equating it with a general logical formula $\phi$.

Other researchers studied the emotional aspect of social norms through psychology experiments or field studies. Kurt et al~\cite{kurt2014behavioral} discuss a study where two subjects try to reach an aggrement on a topic they disagree. The 20 minute scene was split into time intervals were the participant acted in a pro-social, pro-self or neutral way - it was found that these time intervals were distributed according to an exponential distribution, implying that the switching to a new state is some type of non-Markov process. The authors also captured the emotional valence assigned by the participant to different intervals, which was found to follow a random walk pattern.

The impact of emotions and empathy towards a fair or unfair interaction partner was studied by Klimecki et al~\cite{klimecki2016impact}, by introducing the Inequality Game. The social behaviors studied included cooperative or competitive economic choices as well as nice or derogatory verbal behavior.

Several psychological models had been computationally implemented. For instance, Orr et al~\cite{orr2013theory} introduced a dynamic computational model of health behavior (based on a neural network). In their model, the intention is dynamically constructed from both an individual's pre-existing belief structure and the beliefs of others in the individual's social context. This model is based on reconceptualizing the Theory of Reasoned Action~\cite{fishbein1977belief} as a parallel constraint satisfaction system, which allowed the use of a neural network for modeling and prediction.

The difference between public and private behavior, of interest for our model, had been the subject of several research studies. For instance, Houser et al.~\cite{houser2012private} studied the behavior of children in private and public settings in specially created games. The study found that children over age 9, who are typically assumed to exhibit full theory of mind skills, had been better able to resist selfish impulses in public than in private settings, while this difference did not exist for children under 9.

An important aspect of social norms is the degree to which they are actually followed - an aspect often referred to as the ``tightness'' or ``looseness'' of the culture. Gelfand et al.~\cite{Gelfand-2011-TightLooseCultures} show the results of a study that compares the cultural tightness or looseness across 33 cultures. Their findings show that there is a very large variation along this dimension. While~\cite{Gelfand-2011-TightLooseCultures} is primarily concerned about nationwide implications - for instance, whether cultural tightness correlates to historical events such as wars or economic models, such as agriculture or hunting-gathering, our perspective is one of prediction of individual action: a person from a tight culture would be more likely to follow the prescriptive part of a CSSM compared to one from a loose culture.

\subsection*{Social norms in theories of politeness and culture}

A theory of social and cultural behavior strongly related to social norms is  Brown and Levinson's {\em politeness theory}~\cite{Brown-1987-Politeness}, which centers around the maintenance of ``face'' defined as the public self-image of the adult human. More specifically, they define the ``positive face'' that refers to one's self esteem and the ``negative face'' that refers to one's freedom to act. While the Brown-Levinson model aims to discover the culture-independent universals in human communication, it is clear that their definition of politeness does not necessarily match the definition of politeness and indeed the desirable behavior in specific cultures. While indirect speech might help preserve face, there are cultures where direct speech is considered polite and desirable. The interpretation of the Brown-Levinson model in the context of specific cultures is a significant ongoing research topic~\cite{Gu-1990-PolitenessChinese,Matsumoto-1988-PolitenessJapanese}.

The Brown and Levinson model is often interpreted in terms of the work of Paul Grice~\cite{Grice-1975-LogicAndConversation} who formulated the {\em cooperative principle} in conversations. According to the four maxims formulated by Grice, speakers in a collaborative conversation should be truthful, provide an appropriate amount of information (not too much, not too little), be relevant and avoid obscurity of expression. Almost always, the desire to be polite (in the Brown and Levinson definition) and the desire to be cooperative (in the sense of Grice's maxims) are countervailing forces. For instance, the indirect strategy is highly polite, but leads to inefficient communication.

Another influential model that specifically aims to quantitatively measure cultural differences, is the {\em cultural dimensions theory} of Geert Hofstede~\cite{Hofstede-2010-Cultures}. The most recent revision of the theory considers six quantitative cultural dimensions: (1)~power distance, the acceptance of unequal distribution of power, (2)~individualism versus collectivism, (3)~uncertainty avoidance (4)~masculinity versus femininity, a metric measuring the balance between assertiveness and competitiveness versus a focus on cooperation, human relations and quality of life, (5)~long term versus short term orientation and (6)~indulgence versus self-restraint. From the point of view of our model, social metrics can be associated with one or more of these dimensions - for instance dignity has relevance to (1) and (4), while wealth to (5) and (6). Furthermore, Hofstede's analysis shows us that even if two cultures define the same set of social metrics, they might weight these differently in practical behavior.

Relatively few projects attempted to put these theories into implemented software. One such example,  Miller et al. \cite{Miller-2008-IntSystems} describes a software product called the Etiquette Engine that uses the Brown-Levinson politeness model to assess the politeness in interactions involving military personnel of common culture but different rank (such as the interaction between a corporal and a major). In a follow-up work~\cite{Miller-2009-UAHCI} the authors create a more complex model that investigates how culture (as exemplified by Hofstede's cultural factors) as well as politeness levels affect the way in which people react to instructions, commands or requests (``directive compliance'').

Bosse et al.~\cite{Bosse-2008-Damasio} formalizes Damasio's theory of consciousness~\cite{Damasio-1999-MakingOfConsciousness}, where consciousness is built up from the distinct elements of emotion, feeling and core consciousness, the latter being defined as the ``feeling of a feeling''.

Another approach to modeling social-cultural behavior is through computational linguistics. For instance, Bramsen et al.~\cite{Bramsen-2011-PowerRelationships} extract social power relationships  from the language used by the speakers, relying on words and features which can be identified by natural language processing software. Danescu-Niculescu-Mizil et al.~\cite{Danescu-2013-ACL} used computational linguistics to determine the type of politeness strategies (in the Brown-Levinson sense) used by Wikipedia authors in their communication logs.

%
%
\section{A computational model of social norms}
\subsection{The formal model of a social scenario}

In a social scenario, a number of persons participate in a social interaction following a recognizable template. The following definition introduces the components of our formal model.

\medskip
\begin{definition}
We call a {\em social scenario} $\mathfrak{S}$ a tuple $\{\mathbf{A}, \boldsymbol\alpha, \boldsymbol\tau, \mathbf{S}, \mathcal{F}, \mathcal{P}\}$, where:
\begin{itemize}
  \item[\quad]$\mathbf{A} = \{A_1, A_2 \ldots\}$ is a set of {\em actors}. Actors are usually individual humans, although they can also be groups, autonomous robots or software agents.
  \item[\quad]$\boldsymbol\alpha = \{ \alpha_1, \alpha_2 \ldots\}$ is a set of distinct {\em action types}. An {\em action} $a$ is characterized by $a(\alpha, A, x_1 \ldots x_n)$, that is, by the action type, the performing actor and a list of parameters of arbitrary length. We denote with $\mathbf{a}=\{a_1, a_2,\ldots\}$ the (not necessarily discrete) space of all possible actions.
  \item[\quad]$\boldsymbol\tau \subset \boldsymbol\alpha$ is the collection of {\em terminal action types}. A terminal action, for any actor and parametrization, terminates the scenario (moves it to a terminal state).
  \item[\quad]$\mathbf{S} =\{S_1, S_2 \dots\}$ is the (not necessarily discrete) collection of {\em full states of the scenario}.
  \item[\quad]$\mathcal{F}$ is the {\em action impact function} $\mathcal{F}:\mathbf{A} \times \mathbf{S} \times \mathbf{a} \rightarrow \mathbf{S}$. We interpret $S' = \mathcal{F}(A,S,a)$ as the new full state of the system if actor $A$ performs action $a(\alpha, A, x_1 \ldots x_n)$ in state $S$.
  \item[\quad]$\mathcal{P}: \mathbf{A} \times \mathbf{S} \rightarrow \boldsymbol\alpha^{*}$ is the {\em progress function}. We interpret $\mathcal{P}(A,S)=\{ \alpha_{p1},\ldots,\alpha_{pn}\}$ as the set of action types available to actor $A$ in state $S$. If the actor can perform a certain action type, it is free to use an arbitrary parametrization of it. If in a given state no actor can perform any action type, we call it a {\em terminal state}.
\end{itemize}
\end{definition}

While the actions are assumed to always succeed, actions with stochastic outcomes can be modeled through the usual game theory technique of a {\em nature actor} taking an action after the human actor, with the nature actor stochastically either accomplishing or not the intent of the human actor's action.

The progress function $\mathcal{P}: \mathbf{A} \times \mathbf{S} \rightarrow \boldsymbol\alpha^{*}$ had been defined on the full state space of the scenario $\mathbf{S}$. This space state is not necessarily discrete and even when it is, its size increases exponentially with the number of variables describing the state, with the base of the exponent being the number of possible values for each variable. For instance, for the Spanish Steps scenario discussed later in this paper, the number of possible states has a magnitude of $10^{20}$ if the variables are quantized into 10 groups.

In the following, we introduce a structure that helps us analyze scenarios by observing that many human interaction scenarios are {\em progress-segmented}, that is, the full states can be grouped into equivalence classes with regards to the output of the progress function.

\medskip
\begin{definition}
We define $\mathbf{P} = \{ P_1, P_2 \ldots P_n \}$ the collection of a finite number of {\em progress states}. A progress state $P$ is a (not necessarily discrete) collection of full states, such that $S \in P \land S' \in P \Rightarrow \forall A ~~\mathcal{P}(A,S) = \mathcal{P}(A,S')$. The {\em progress state discretization} function $PSD:\mathbf{S} \rightarrow \mathbf{P}$ maps states to progress states. We will call the function $\mathcal{P}_R:\mathbf{P} \times \mathbf{A}\rightarrow \boldsymbol\alpha^{*}$ the {\em reduced progress function} and define it as $\mathcal{P}(A, S) = \mathcal{P}_R(A, PSD(S))$.
\end{definition}

Progress states represent a reduction of the full state space because each progress state corresponds to multiple full states. For instance, in a negotiation scenario, all the states where a deal was reached will be grouped into the same progress state. There is always a discrete number of progress states while the full state space can be continuous. Thus, in contrast to $\mathcal{P}$, the reduced progress function $\mathcal{P}_R$ is defined on a discrete and (usually) small space.

A scenario $\mathfrak{S}$ is defined generically over actors: for instance, in a commercial transaction there is a buyer and a seller actor. When the scenario actually takes place, the buyer and the seller will be played by specific  social agents, let us say Jack and Jill. The same social agent might be part of multiple scenarios: for instance Jack might be a seller in one scenario instance and the buyer in another. Social agents maintain their own {\em private state} $S_\textit{SA}$. The state of a given scenario is a superset of the union of the private states of the social agents playing the actors. T

The private state of a human social agent includes includes her memories, as well as the full cognitive, emotional and physiological state~\cite{Danziger-2011-ExtraneousFactors}. While this is clearly beyond our modeling capacity, we also notice that in most situations we expect our social partners to follow the social norms and restrain the impact of basic emotions and physiology on their behavior. Thus, for our model we only consider the subset of the private state of the social agent composed of (a) the {\em culture-sanctioned social metrics} (b) a small set of {\em beliefs about concrete facts} relevant to the current scenario.

%
%
\subsection{Culture-sanctioned social metrics}

We define a culture-sanctioned social metric (CSSM) as the degree to which a social value considered desirable by given culture is upheld in a given social interaction. We say that a culture {\em sanctions} a metric if (a) provides a {\em name} for it (b) provides an (informal) {\em algorithm} for its evaluation (c) expects its members to {\em continuously evaluate} the metric for themselves and salient persons in their environment
and (d) provides {\em rules of conduct} which depend on the metric. Seen like this, CSSMs are numerical stand-ins for a collection of related social norms.

The CSSMs can be either tangible or intangible. Tangible metrics such as financial wealth or time spent doing something can be measured by physical means (although many times they are only estimated). Intangible metrics, such as politeness, dignity, hospitality, generosity, piety, ``face'' or ``manliness'' are socially constructed, not directly measurable and depend on the specific culture. The separation between tangible and intangible metrics is often fuzzy, because even many tangible metrics, such as time, are subject to interpretation by the human agents.

Loosely speaking, we can say that social agents try to maximize the CSSMs they subscribe to. Nevertheless, actions might have opposing effects on CSSMs ({\em e.g.} actions that optimize generosity decrease financial wealth).

To be educated in a culture an individual must know not only the name of the metric, but also the evaluation algorithm and the rules of conduct associated with it. It is not guaranteed that a given individual will {\em follow} the rules of conduct - however, he or she will be {\em aware} of the rules and their transgression.

The same name might define different metrics in different cultures. For instance, the word ``dignity'' has different evaluations and rules of conduct in different English speaking cultures. The dictionary translation of the word in other languages, such as ``azmat'' in Urdu, ``pratistha'' in Hindi or ``m\'elt\'os\'ag'' in Hungarian, can denote even more divergent CSSMs. This being said, there are many CSSMs that appear in several cultures in identical or near-identical form. There are groups of cultures with closely related metrics - for instance the cultures aligned with the Western European model, the culture of China and nations influenced by Chinese culture and the cultures of the Near East and North Africa. In addition, certain CSSMs are cross-cutting geographical, language and religious boundaries, such as the striking similarities between ``cultures of honour'' in places as far away as the Scottish highlands, the Bedouins of the Sahara or the Southern USA~\cite{Nisbett-1996-CultureOfHonor}.

Many rules of conduct associated with CSSMs consider not only the actor's own perspective, but also the perspective of other actors in the scenario. For instance, gestures of politeness and respect are often enacted such that they are visible to and noted by not only the direct interaction partner, but also by third parties. Taking this into consideration, we propose a model where a specific CSSM is identified by five parameters: {\tt \small CSSM(C,M,SA,PA,EA)}, where:

\begin{itemize}
  \item {\tt C} is the culture that defines the CSSM and specifies its rules.
  \item {\tt M} is the name of the metric, which is unique in the given culture (but different cultures might mean different metrics under the same name).
  \item {\tt SA} is the {\em subject agent} characterized by the metric.
  \item {\tt PA} is the {\em perspective agent}, from whose perspective the metric is evaluated.
  \item {\tt EA} is the {\em estimator agent}, who estimates the CSSM.
\end{itemize}

The intuition about the three agents is as follows: in the estimation of {\tt EA}, the agent {\tt PA} believes that the value of the metric {\tt M} for agent {\tt SA} is equal to {\tt \small CSSM(C,M,SA,PA,EA)}. There is no requirement for {\tt SA}, {\tt PA} and {\tt EA} to be all different. For a CSSM to play a role in a scenario, we need {\tt EA} to be cognizant of culture {\tt C}. In addition, it is necessary for {\tt EA} to believe that {\tt PA} is cognizant of culture {\tt C} (although this belief might be incorrect). It is not necessary for {\tt SA} to be cognizant of the culture (although whether he is or not might be a factor in the behavior of other actors). A specific CSSM is always part of the {\em private state of the estimator agent} $S_\texttt{EA}$.

Let us consider several examples. {\tt \small CSSM(Western,dignity,John,John,John)} represents John's estimate of his own dignity, in the Western cultural model. {\tt \small CSSM(Western, politeness,John,Mary,John)} represents John's estimate about how Mary sees his politeness. If John cares about Mary's opinion, he will adjust his behavior in such a way that Mary's perspective will improve. Note that this value might not be identical to {\tt \small CSSM(Western,politeness,John,Mary,Mary)}, that is, Mary's own opinion about John's politeness.

For a case of cross-cultural perspective let us consider the case of J{\'a}nos, a Hungarian businessman in China, who publicly admits to a business partner Chen a mistake in formulating a purchase order. This will affect {\tt \small CSSM(Chinese,Face,J{\'a}nos,J{\'a}nos,Chen)} that is, Chen's estimate of  J{\'a}nos's own estimate of loosing face. In this context, Chen might not understand why J{\'a}nos would do such a thing. What happens here, is that Chen is evaluating a CSSM which J{\'a}nos does not: J{\'a}nos is not educated in Chinese culture, and the concept of ``face'' as a metric is not sanctioned in Hungarian culture. Thus {\tt \small CSSM(Hungarian,Face,J{\'a}nos,J{\'a}nos,J{\'a}nos)} is not defined, while {\tt \small CSSM(Chinese, Face,J{\'a}nos,J{\'a}nos,J{\'a}nos)}, while defined, it cannot be evaluated by J{\'a}nos, who does not know the Chinese culture. Nevertheless, this CSSM {\em can} impact the outcome of the scenario: for instance, Chen might act to prevent J{\'a}nos from loosing face, even if J{\'a}nos is unaware of this.

\subsection{Concrete beliefs}

The evaluation of a CSSM is often conditional on aspects of the current scenario -- an action might break a social norm in some situations, but be perfectly acceptable in others. Thus, the behavior of social actors also depends on their beliefs about certain aspects of the current scenario. For CSSMs where the subject, perspective and estimator agents are not the same, this also means conditioning on beliefs about beliefs of other agents, effectively the theory of mind deployed by the social agent. Although reasoning about {\em general} human beliefs is notoriously difficult, fortunately, social norms usually depend only on the circumstances of the scenario, not on the ``deep beliefs'' of the participants. We will thus focus on a very restricted set of beliefs that pertain to simple binary questions that can be, in principle, unequivocally answered by an omniscient external observer. Such {\em concrete questions}  include: ``Is A holding a flower?'' or ``Are A and B engaged in a commercial transaction?''.

In contrast to the omniscient external observer, the actors in the scenario need to work with incomplete knowledge and limited rationality. We will call {\em concrete beliefs} (CBs) the beliefs maintained by the actors in a scenario with regards to the answers of concrete questions. We say that a scenario {\em defines} a CB if (a) there is an algorithm that an omniscient external observer could use to unequivocally answer the question underlying the CB (b) the scenario expects at least one actor to {\em continuously evaluate} the CB for himself and other salient actors in the scenario and (c) the scenario provides {\em rules of conduct} that depend on the CB {\em or} the CB affects the calculation of CSSMs.

The definition of CBs has clear analogies to the definition of CSSMs, but several important differences exist. First, CBs do not depend on the culture: while the definition of politeness varies from culture to culture, the question whether a person holds a flower or not is decidable without cultural references. Instead of being tied to the culture, the CBs are tied to a specific scenario. Another difference is that while CSSMs represent the social values of a subject actor, {\em e.g.} the politeness of John, the concrete question can refer to any aspect of the scenario, including inanimate entities (``is it raining?'').

Putting these considerations together, we will identify a concrete belief with four parameters: {\tt CB(SC,BD,PA,EA)}, where:

\begin{itemize}
  \item {\tt SC} is the scenario instance that specifies the question.
  \item {\tt BD} is the description of the belief (normally, through the associated question).
  \item {\tt PA} is the {\em perspective actor}, from whose perspective the belief is evaluated.
  \item {\tt EA} is the {\em estimator actor}, who performs the estimate and owns the knowledge.
\end{itemize}

A number of considerations discussed in the case of CSSMs are applicable to CBs as well. The CB is always part of the {\em private state of the estimator actor} $S_\texttt{EA}$. Although there is a requirement for some actors to evaluate specific CBs, this evaluation might be incomplete or incorrect due to the lack of information, misunderstanding or cognitive overload.

%
%
\section{Tracking CSSMs and CBs in a social scenario}

The action impact function (AIF) $\mathcal{F}:\mathbf{A} \times \mathbf{S} \times \mathbf{a} \rightarrow \mathbf{S}$ describes the way in which the state of a scenario instance evolves under the impact of a specific action performed by an actor: $S' = \mathcal{F}(A,S,a)$ is the new state of the system if actor $A$ performs action $a(\alpha, A, x_1 \ldots x_n)$ in state $S$.

Taking advantage of the fact that we reduced the state to a collection of CSSMs and CBs $S = \{\textit{CSSM}_1, \ldots, \textit{CSSM}_n, \textit{CB}_1, \ldots, \textit{CB}_m\}$ allows us to split the AIF into a collection of functions, one for each component of the state:

\begin{equation}
\left\{
\begin{array}{rcl}
\textit{CSSM}_i' & =  & F_i^\textit{CSSM}(A, S, a) \\
\textit{CB}_j' & = & F_j^\text{CB}(A, S, a)
\end{array}
\right.
\end{equation}

In order to model concrete examples, we will need to choose mathematical representations for the CSSM and CB AIFs. In the following two sections we describe one particular choice that worked for us given the scenario we aimed to model and the data we had available: use sums of products of logistic functions for the CSSM AIFs and the Dempster-Shafer theory of evidence for the CB updates. Naturally, other choices exist, and might be more suitable under different circumstances. For instance, one could use deep neural networks for the CSSM AIF - if there is sufficient data to train them. For the CB AIF, another choice would be a hidden Markov model or a dynamic Bayes network. Recurrent neural networks (LSTM or GRU based) are another possibility for the CB maintenance, if sufficient training data is available.

%
%
\subsection{CSSM AIFs using sums of products of logistic functions}
\label{sec:ActionImpactFunctions-CSSM}

To create a CSSM AIF we must decide on (a) the subset of the state relevant for a given CSSM, (b) the shapes of the AIFs and (c) the parametrized mathematical forms that can represent these shapes in a convenient way. We have seen that the state S is composed of the private states of the participating agents. The update of a CSSM in the form of {\tt CSSM$^i$ = CSSM(C,M,SA,PA,EA$_x$)} will be kept and maintained by the estimator actor {\tt EA$_x$}, and this actor only has access to the other CSSMs and CBs {\em in its own private state}: {\tt CSSM(\_,\_,\_,\_,EA$_x$)} and {\tt CB(\_,\_,\_,EA$_x$)}. Thus, $F_i^\textit{CSSM}$ will be a numerical function depending only on the CSSMs and CBs whose estimator agent is the same as the estimator agent of {\tt CSSM$^i$}.

Let us now discuss the shapes an AIF would likely take. Tangible CSSMs like time or money, usually have simple AIFs. For instance, if an action takes time $t_a$ then the action will add this value to the ``time'' CSSM. If the action involves paying the sum of $m_a$ dollars, this will decrease the ``wealth'' CSSM with the given value.

Things are more complicated for intangible metrics, where the change can be nonlinearly dependent on multiple factors. For instance, for the social metric of dignity, many cultures have a {\em sensibility threshold}: they advise to ignore trifling offenses. Similarly, there is a {\em saturation threshold}: a level at which the offense is so big that further increasing it would not affect the dignity level. Thus, typical AIF shapes might contain positive or negative slopes, thresholds and saturation behaviors, but it will unlikely to involve periodic functions or multiple local maxima. Furthermore, the change in social metrics often depends on the beliefs: we are less offended by the angry voice of the interaction partner if we believe that her anger is justified.

There are many kinds of mathematical expressions that can generate such shapes. Our goal is to balance computational and modeling convenience with the hope of capturing some of the essential phenomena behind the metrics. Many metrics closely related to CSSMs are modeled in psychology with the assumption of certain consumable resources in the human psyche (see for instance the hypothesis of ``ego depletion'' \cite{Baumeister-1998-Ego}). In some cases, these consumables can be actually identified as physiological measures such as the blood glucose level~\cite{Gailliot-2007-Physiology}. Hagger et al~\cite{hagger2009strength} conceptualizes self-control as a limited resource, whose depletion, for instance, would reduce the ability to follow dieting rules. Values dependent on depletable resources are often modeled using the sigmoid shaped logistic curve $f(x)=1 / (1 - e^{-x})$. To allow for a more flexible representation, we start with a version of Richard's curve~\cite{Richards-1959-Flexible}, a logistic function parametrized with six intuitive parameters in the form:
\begin{equation}
Y(t) = A + \frac{K-A}{\left(1+Qe^{-B(t-M)}\right)^{1/v}}
\end{equation}
In this formula, $A$ is the lower asymptote, $K$ the upper asymptote, $B$ the growth rate, while $v$, $Q$ and $M$ are parameters which affect the location and rate of maximum growth of the function. The six parameters allow for considerable freedom in the specification of the shape of the sigmoid function, but they also provide more detail than the requirements of our problem domain. Thus, we chose to reduce the number of parameters by only keeping as variables $K$ for the upper asymptote, $M$ for the location of largest growth and $B$ for the growth rate. The other values will be fixed at $A=0$ and $Q=v=1$. The intuitive nature of these parameters allow us to handcraft appropriate function shapes. We will call this 4-parameter function the {\em logistic component} of the AIF:
\begin{equation}
L(x, K, M, B) = \frac{K}{1+e^{-B(x-M)}}
\end{equation}

\begin{figure}
\centering
\begin{tabular}{cc}
\includegraphics[width=0.45\columnwidth]{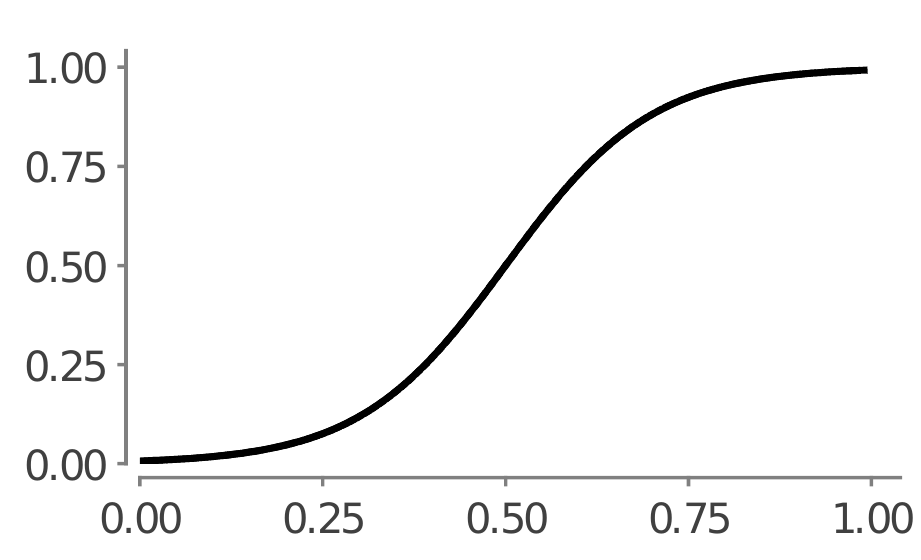} &
\includegraphics[width=0.45\columnwidth]{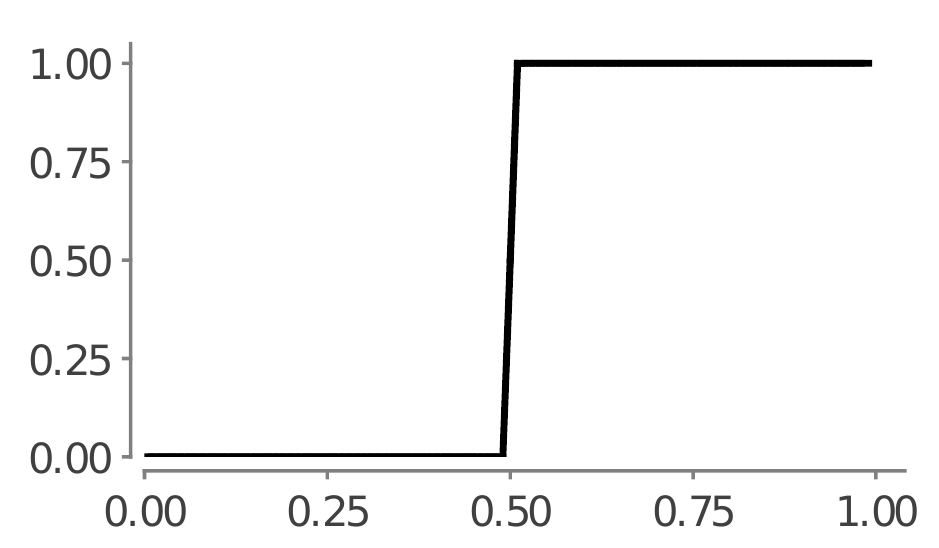} \\
(a) & (b) \\
\includegraphics[width=0.45\columnwidth]{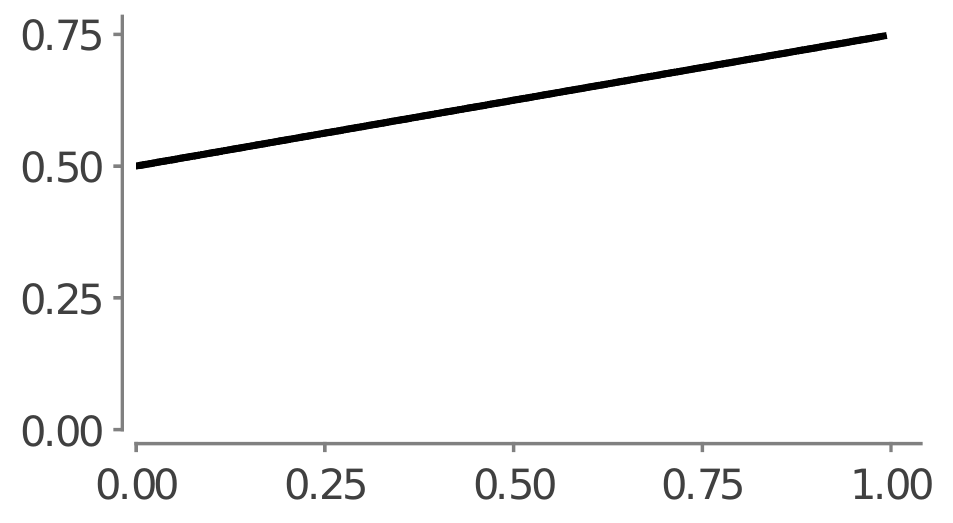} &
\includegraphics[width=0.45\columnwidth]{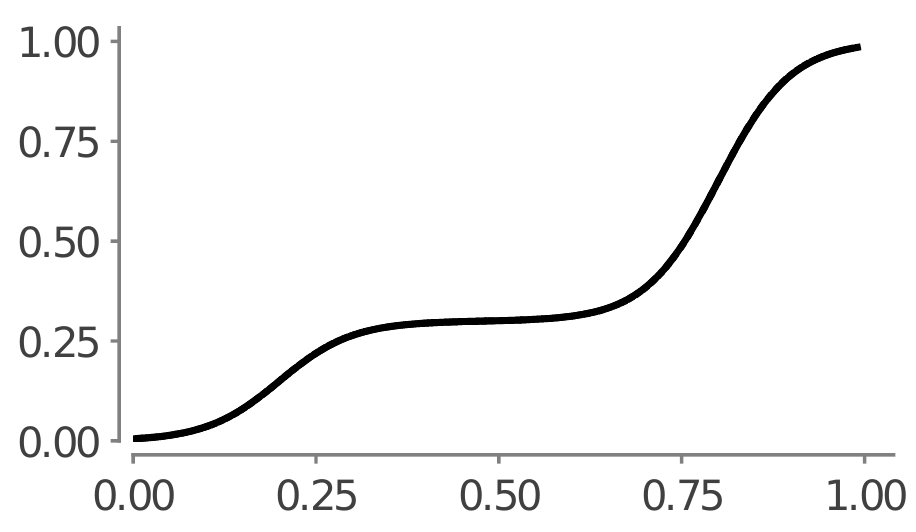} \\
(c) & (d) \\
\end{tabular}
\caption{{\bf Implementing different AIF functions shapes using sums of logistic components.} (a)~Sigmoid shape L(x,1,0.5,10) (b)~Step-function shape L(x,1,0.5,1000) (c)~Linear growth shape L(x,5,0,0.2) + L(x,-2,-10,100) (d)~Multi-plateau shape L(x,0.3,0.2,20) + L(x,0.7,0.8,20)}
\label{fig:AIFLogistic}
\end{figure}

Thus, our representation for the CSSM AIFS will be a sum of products of logistic functions:

\begin{equation}
F_i^\textit{CSSM} = \sum_{k} \Big( \prod_l  L(x_{kl}, K_{kl}, M_{kl}, B_{kl}) \Big) \label{eq:aifcssm}
\end{equation}
\noindent where $x_{kl}$ is either the constant 1, an arbitrary parameter of the action, a CSSM or a CB. The CSSMs and CBs appearing in the formula must have the same estimator as CSSM$^i$. When some of the logistic components recur in more than one term, we will sometimes write an AIF more compactly by factoring them out. In practice, we found that we need a very small number of logistic components to achieve the desired shapes -- usually only one or two components per dimension. Figure~\ref{fig:AIFLogistic} shows four examples of such function shapes achieved with at most two logistic component terms: a sigmoid shape, a step function shape, a linear slope and a multi-plateau shape with two saturation plateaus.

\subsection{CB AIFs using the Dempster-Shafer theory of evidence}
\label{sec:ActionImpactFunctions-CB}

CB AIFs update the beliefs held by the evaluator actor as a result of an action. The Dempster-Shafer~\cite{Shafer-1976-TheoryOfEvidence,Yager-1987-Dempster} theory of evidence represents belief and confidence levels in a single computational model. The belief supported by the accumulated evidence is represented by a {\em mass function} that assigns fractions of a mass of 1.0 to all non-empty combinations of beliefs. New evidence changes the distribution of the mass. Using the mass function, the belief in a statement can be calculated as a value bounded by two intervals, the {\em belief} (or {\em support}) and the {\em plausibility}. The difference between these two values represent the uncertainty associated with the belief. In our implementation, the full mass functions are part of the private state of the estimator agent $S_\texttt{EA}$, however, the CSSM calculations equate the CB with the support component of the Dempster-Shafer model.

Let us consider the concrete question ``Are A and B engaged in a commercial transaction?'' and let us assume that the current CB of an agent is represented by the mass distribution $m(\texttt{true})=0.4$, $m(\texttt{false})=0.1$ and $m(\texttt{true,false})=0.5$. With these settings the Dempster-Shafer values will be $\textit{support}(\texttt{true})=0.4$ and $\textit{plausibility}(\texttt{true})=0.9$, which implies the CB to be $0.4$.

As the concrete questions underlying the CBs are binary, the CB can be characterized by the mass function values for $m(\texttt{true})$ and $m(\texttt{false})$. As $m(\emptyset)=0$ by definition, we will have $m(\texttt{true}\textit{ or }\texttt{true}) = 1 -m(\texttt{true}) - m(\texttt{false})$.

An evidence arriving in the form of new information received from an action performed by an agent other than the estimator\footnote{Actions taken by the estimator agent itself will never impact its own CBs. As the agent is free to choose its own action, the choice of the action never represents new information.} or an event will also have a mass function $m_e$. The new belief value will be given by Dempster's rule of combination (the {\em conjunctive merge}):
\begin{equation}
m'(A) = \frac{1}{1-K}\sum_{B \cap C=A\neq\emptyset} m(B)\cdot m_e(C)
\end{equation}
\noindent where $A, B, C \in \{\texttt{true}, \texttt{false}, (\texttt{true}\textit{ or }\texttt{false})\}$ and
\begin{equation}
K = \sum_{B \cap C=\emptyset} m(B)\cdot m_e(C)
\end{equation}
A special consideration must apply to CBs where the perspective or estimator agent is a group agent, such as a crowd. Naturally, different members of the crowd can hold different beliefs. One natural way to model this is to consider that each of the members contribute to the overall mass function with a fractional mass. For instance, for a crowd of 100 people, each of them will have a personal mass function where the masses add up to 0.01. For the group agent representing the crowd, the masses of different beliefs will be the sum of the individual masses held by the members.

%
%
\section{Validating the model: the Spanish Steps flower selling scam}

The main goal of our proposed computational model of social norms is to support software implementations for the modeling and simulation of social systems as well as to generate behaviors in human-robot or human-agent systems. The formal model is described in the previous sections, and a software implementation is available at {\tt https://github.com/NetMoc/CSSM}. But the existence of a formalism and a software implementation does not guarantee the usefulness of the model. One way to validate our approach is to show that we can gain non-trivial insights. In this section we show how our model can explain apparently irrational human behavior in a complex social scenario.

Let us consider a flower selling scam perpetrated by crooked sellers at the Spanish Steps in Rome (and probably at many other popular destinations around the world). The goal of the seller is to pressure a client (typically a woman or a romantic couple) to purchase a flower at an inflated price by proceeding through the following scenario:

\begin{itemize}

\item The seller offers a bouquet of flowers to the client. The client declines to purchase.

\item The seller offers a single flower, relying on gestures implying that it is a gift. If the client refuses to take the flower, he repeats the offer several times, pushes the flower into the client's hands, or inserts it into her bag.

\item The seller waits an amount of time at some distance from the client. During this time, the client gets used to the received gift, takes a picture with it or puts it in her bag.

\item The seller approaches the client and requests payment, relying on visual signals (rubbing the pointing finger and thumb together).

\item The client repeatedly attempts to return the flower while the seller refuses to take it. The action concludes by either the client paying or by escalating her verbal efforts to return the flower until the seller decides to take it back.

\end{itemize}

The Spanish Steps flower scam, despite being physically simple, is based on a series of complex decisions. It is, at its roots, a negotiated commercial transaction, which, however, is initiated by a {\em deceit} -- the implication that the flower is a gift. The deceit is facilitated by the {\em blocking of the normal channels of communication} -- the seller is usually a good speaker of several languages, but fakes reduced communication ability to position the deceit as a misunderstanding. The successful conclusion of the scam relies on the {\em manipulation of the public perception:} the client needs to have the impression that everybody around believes that she agreed to buy the flower.

Explaining and predicting the behavior of the participants is not necessarily easy even for the human observer. Why some clients accept to pay for the flower, well knowing that they are cheated? Conversely, why does the seller, occasionally, give up, without pushing the selling process to the extremes? Neither question can be answered based on the assumption of a narrowly defined wealth-maximizing rational agent. The intuitive answer is that a successful seller manipulates the client to believe that not purchasing the flower would violate social norms, while a successful client manipulates the seller to a situation that further pushing the sell would violate its own social norms. In the following we model this scenario using our framework to put this intuition on a quantitative basis.

%
%
\subsection{Actors, progress graph and action types}

The Spanish Steps scenario has three individual actors: the {\tt \small Seller}, {\tt \small Client}, {\tt \small Spouse} and a group actor, the {\tt \small Crowd}. Only the {\tt \small Seller} and the {\tt \small Client} take actual actions; the {\tt \small  Spouse} and the {\tt \small Crowd} influence the outcome by being the perspective actors in CSSMs and CBs considered by the active actors.

\begin{figure}
\centering
\includegraphics[width=\columnwidth]{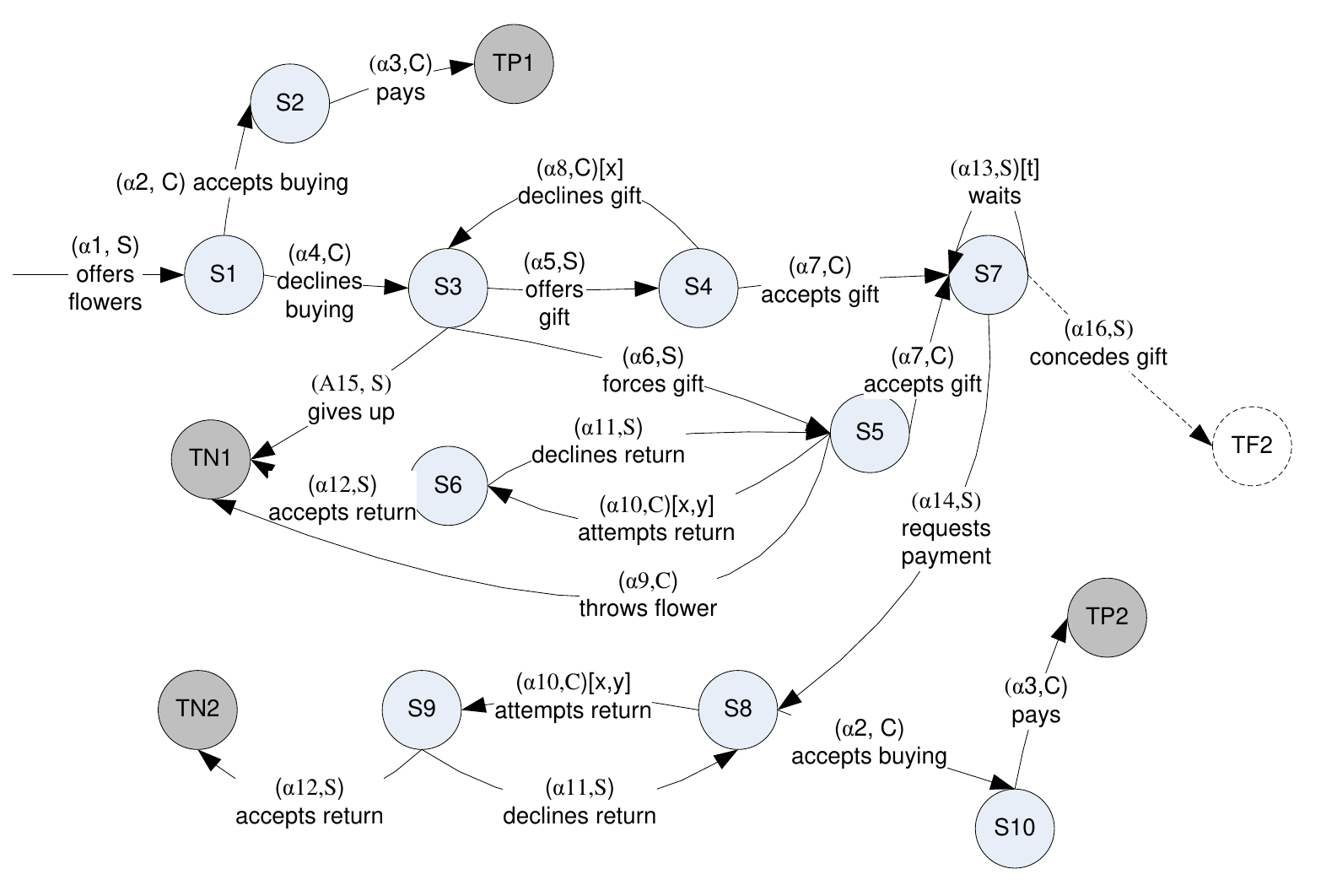}
\if notdefined
{
\footnotesize
\begin{tabular}{|l|l|}
\hline
Action type & Description \\
\hline
$\alpha 1$ & offers flowers to sell \\
$\alpha 2$ & accepts to buy the flowers \\
$\alpha 3$ & pays for flowers \\
$\alpha 4$ & declines buying flowers\\
$\alpha 5$ & offers flower as gift\\
$\alpha 6$ & forces gift\\
$\alpha 7$ & accepts flower as gift\\
$\alpha 8$ & declines gift (x,y)\\
$\alpha 9$ & throws gift\\
$\alpha 10$ & attempts return of flower (x,y)\\
$\alpha 11$ & declines return of flower\\
$\alpha 12$ & accepts return of flower\\
$\alpha 13$ & waits (t)\\
$\alpha 14$ & requests payment\\
$\alpha 15$ & gives up interaction\\
$\alpha 16$ & concedes gift\\
\hline
\end{tabular}
}
\fi
\caption{\label{fig:ProgressGraphSpanishSteps}
{\bf The progress graph of the Spanish Steps scenario.}}
\end{figure}

The two active actors can take actions belonging to the 16 action types listed in Figure~\ref{fig:ProgressGraphSpanishSteps}. For some of these actions we also need to consider the parametrization. $\alpha 8$ and $\alpha 10$ are actions involving verbally and gesturally declining a gift and attempting to return the flower respectively. They are parametrized by their ``loudness'' $x$ that determines how many onlookers will overhear the transaction and their ``offensiveness'' $y$ that influences how the action impacts the politeness of the actor and the dignity of the target. For action type $\alpha 13$, which involves the {\tt \small Seller} waiting without taking any action, the parameter is the length of the wait $t$. The Spanish Steps scenario can be represented with the progress state graph shown in Figure~\ref{fig:ProgressGraphSpanishSteps}, which has 10 non-terminal progress states and 4 terminal progress states.

The full state space of the scenario, depending on the degree at which we choose to model the state of the client and the seller, can be very large. The progress state discretization function however, groups these states in the 14 progress states. For instance, progress state S9 represents a situation where the client had just tried to return the flower. This progress state groups a large number of possible full states - from states in which the client is mildly amused to states in which she feels angry, humiliated, embarrassed at various degrees, as well as possible combinations of these.

%
%
\subsection{CSSMs in the Spanish Steps scenario}

The metric of financial wealth is the central concern of every financial transaction. However, as we have seen, taken by itself, the assumption of maximizing financial wealth cannot explain or predict the behavior of the actors in the Spanish Steps scenario. In the following, we will consider a collection of CSSMs and CBs that allow us to model the scenario with explanatory and predictive power.

We will use four CSSMs: two concrete ones ({\tt \small Wealth} and {\tt \small Time}) and two intangibles ({\tt \small Dignity} and {\tt \small Politeness}), all of them defined in the Western culture.

\bigskip

{\tt \small Wealth:} is the sum of the financial wealth of the person, measured in real-world currency. The social norms associated with wealth in the western culture encourage users to increase their own personal wealth. For the Spanish Steps scenario we assume that the seller and the client each consider only their own personal wealth\footnote{There are scenarios where the estimation of the wealth of the interaction partner is necessary for accurate modeling: for instance, in a ``Giving Money to a Beggar'' scenario.}:

\smallskip

\noindent {\tt \small CSSM(Western,Wealth,Seller,Seller,Seller)}

\noindent {\tt \small CSSM(Western,Wealth,Client,Client,Client)}

\bigskip

{\tt \small Time:} is the amount of time spent in the current scenario measured in seconds. Western social norms discourage people from wasting their time (``Time is money'')\footnote{As we described CSSMs as values to be maximized, technically the CSSM would be ``saved time'' not ``elapsed time''.}.  Again, we assume that the seller and the client only consider their own time spent. If the seller deals with one client at a time, these values will be the same:

\smallskip

\noindent {\tt \small CSSM(Western,Time,Seller,Seller,Seller) = }

\noindent {\tt \small CSSM(Western,Time,Client,Client,Client)}

\bigskip

{\tt \small Dignity:} in Western culture is associated with the degree of respect the person receives from interaction partners or the degree of self-respect he shows. Dignity being an intangible metric, the rules associated with it are more complex. An insult decreases the dignity of a person. A person will feel insulted if the communication partner uses rude language, or if he perceives that he has been lied to. The dignity of a person is also affected by his own actions: for instance, an excessive emotional display decreases the dignity of the person. It is considered undignified to renege on a promise (for instance, to not fulfill an accepted commercial transaction).

As Western culture requires persons to maintain their dignity, the metric affects the decisions of the actors in the scenario. The client evaluates his own dignity from his own perspective, from the perspective of the spouse and the perspective of the crowd. These values are also estimated by the seller. As the seller sees all the relevant actions and understands Western culture, his estimate of the client's dignity will be the same as the client's own estimate.

Modeling the dignity of the seller presents an interesting challenge. We might say that the seller, engaged in a deceitful selling maneuver, does not care about his own dignity or at least values it much less than financial gain\footnote{An alternative explanation would be that the seller has a different culture and thus applies a different metric.}. Nevertheless, even if he does not care about his own dignity in the Western definition of the metric, social pressure obliges him to consider his dignity from the perspective of the crowd. This value can also be estimated by the client:

\smallskip

\noindent {\tt \small CSSM(Western,Dignity,Seller,Crowd,Seller)} =

{\tt \small CSSM(Western,Dignity,Seller,Crowd,Client)}
\bigskip

{\tt \small Politeness:} for the purposes of this paper, we define the Politeness CSSM as the degree to which a social actor adheres to acceptable forms of speech and gesture in specific circumstances\footnote{We are considering here a relatively narrow definition of the politeness of speech forms and gestures. This is a more restricted and specific interpretation than, for instance, positive face in politeness theory~\cite{Brown-1987-Politeness} which tries to account for a wider range of phenomena across cultures.}. Persons are required to maintain a positive politeness in the perception of the self, as well as from the perspective of peers and crowd. Due to the low power distance in Western culture the rules associated with politeness are comparatively simple, but by no means trivial. A person decreases his politeness metric if he uses rude language, loud voice or indecent or threatening gestures. It is considered impolite to decline a gift or to insist on an issue in the face of the refusal from the interaction partner. In contrast to cultures that insist on politeness under any circumstances, politeness rules in Western culture take into account whether the interaction partner ``deserves'' politeness based on his recent actions. For instance, rude language addressed to a crooked seller has a smaller impact on the client's politeness metric.

The politeness metric is taken into account at several action choices. The client's decision to accept the flower in state S4 is influenced by his self perception:

\smallskip
\noindent {\tt \small CSSM(Western,Politeness,Client,Client,Client)}
\smallskip

At progress states S8 and S9 however, the client knows that he is being cheated, so his rudeness towards the obviously crooked seller will not affect his own politeness metric. However, he still needs to worry about the perception of the crowd and his spouse who might not consider the seller crooked:

\smallskip
\noindent {\tt \small CSSM(Western,Politeness,Client,Crowd,Client)}

\noindent {\tt \small CSSM(Western,Politeness,Client,Spouse,Client)}
\smallskip

The metric of politeness is also relevant to the seller, who must care about his own politeness as perceived by the crowd:

\smallskip
\noindent {\tt \small CSSM(Western,Politeness,Seller,Crowd,Seller)}
\smallskip

This fact is also known by the client, who can approximate this value with:

\smallskip
\noindent {\tt \small CSSM(Western,Politeness,Seller,Crowd,Client)}
\smallskip

%
%
\subsection*{CBs in the Spanish Steps scenario}

The next step is to determine the concrete questions and associated CBs that influence the behavior of the actors in the Spanish Steps scenario. We find that we only need to consider two questions: {\tt \small Q-Gift} and {\tt \small Q-Agreed}.

\bigskip

\noindent {\tt \small Q-Gift: Is the flower a gift?}

\smallskip

This question is unequivocally answerable by the seller (he knows it is not) so we have:

\smallskip
{\tt \small CB(S,Q-Gift,Seller,Seller) = 0}
\smallskip

However, the value for the client {\tt \small CB(S,Q-Gift,Client, Client)} has a significant impact on whether he will accept the flower or not in progress state S3. If the flower is a gift and he declines it, the client will incur a penalty in politeness. On the other hand, it is not impolite to decline a commercial transaction. Accordingly, the seller is interested to ensure that in state S3 we have a high value for {\tt \small CB(S,Q-Gift,Client, Seller)}, a value that approximates the client's own belief.  The value {\tt \small CB(S,Q-Gift,Client,Client)} will be set to 0 at the moment when the seller asks for money.

\medskip

Let us now move on to the second relevant question:

\smallskip
\noindent {\tt \small Q-Agreed: Has a commercial transaction been agreed upon?}
\smallskip

The answer to this question is actually clear for both the self and reciprocal
CBs of the client and the seller, as they both know that no commercial
transaction took place\footnote{As a note, one could imagine a scenario where
the client might be confused whether he had actually agreed to a transaction
without really noticing it. However, this would not normally happen in this
scenario: the client knows that he is being cheated.}. Thus {\tt \small
CB(S,Q-Agreed,X,Y) = 0} for all cases where {\tt \small X} and  {\tt \small Y}
are the  {\tt \small Seller} or  {\tt \small Client}.

The interesting CB in this case is {\tt \small CB(S,Q-Agreed,Crowd, Client)}. If
this value is high, the action $\alpha 10$ is perceived by the crowd as reneging
an agreed upon transaction, while if it is low, they judge it to be a
confrontation with a crooked seller, with different impacts on the politeness
and dignity CSSMs. The seller must thus act to bring {\tt \small
CB(S,Q-Agreed,Crowd,Client)} $\approx$ {\tt \small CB(S,Q-Agreed,Crowd,Seller)}
to a high value.

Notice the importance of a passive actor (the crowd) in the scenario. The Spanish Steps scam would rarely succeed on an empty street. The presence of the crowd, even without taking any active action, changes the dynamics of the scenario by serving as a perspective actor for the dignity and politeness CSSMs and the {\tt \small CB(S,Q-Agreed,..,..)} values.

%
%
\subsection{CSSM AIFs in the Spanish Steps scenario}

In the following we illustrate some of the representative CSSM AIFs, moving from simpler to more complex. These AIFs had been knowledge engineered as follows. We started with an informal description of the impact in natural language. Then, we separated the parameters of the action and identified the ways in which they  change the CSSM (step functions, linear dependency, single or multiple plateaus). For each of these dependencies we chose appropriate logistic components as seen in Figure~\ref{fig:AIFLogistic}, and adjusted the parametrization using grid search until it matched the natural language description. Finally, we combined the components to obtain the appropriate multi-variable AIFs. Considerations of space prevent us from presenting this process in detail for the individual AIFs.

%
%
\subsubsection{Wealth of the client at $\alpha 3$ (client paying for the flower)}

We assume that the cost of the flower is 5\texteuro ~(we do not model bargaining for the price). Thus, if we denote with $v=${\tt\small~CSSM(Western,Wealth,Client,Client,Client)} the value before the action, and with $v'$ the same value after the action we have $v'=v-5$. Using logistic components this can be approximated by:
\begin{equation}
v' = L(v, 50, 0, 0.08) + L(v, -30, -100, 100)
\end{equation}

%
%
\subsubsection{Time of crowd at $\alpha 13$ (wait time t before asking for money)}

Naturally, the time passes the same way for all the actors, independently of perspective. The CSSM of interest is
$v=$ {\tt \small CSSM(Western,Time,Crowd,Client,Seller)}= {\tt \small CSSM(Western,Time,Crowd,Client,Client)} because this is the value that impacts the evolution of the belief of the crowd that a commercial transaction had been agreed upon {\tt \small CB(S,Q-Agreed,Crowd,Client)}.

In this case, we simply add the parameter to the time value $v'=v+t$, approximated by the following LCF expression:
\begin{equation}
v' = L(v, 50, 0, 0.08) + L(v, -25 + t, -100, 100)
\end{equation}

%
%
\subsubsection{Impact of $\alpha 10$ on the estimated public perception of client politeness}

Action $\alpha 10$ represents the attempt to return
the flower. The action is parametrized by the parameters $x$ (loudness) and $y$
(rudeness). We calibrate the numerical values of these parameters on the scale
of [0,1] using the keywords in  Table~\ref{tab:RudenessLoudness}. We are using common sense values for the loudness\footnote{Note that, strictly speaking, the loudness can be matched to physically measurable sound pressure values, but this is less useful in developing the AIF than the intuitive metrics used here.}. The low values of the rudeness parameter (0.0-0.6) are mapped to the mitigation level of speech. Higher values of rudeness involve insulting language and threats of physical violence.

\begin{table}
\caption{Intuitive keywords for calibrating the parameters of action $\alpha 10$}
\label{tab:RudenessLoudness}
\begin{tabular}{|p{1cm}|p{2.5cm}|p{3.9cm}|}
\hline
Value & $x$ (loudness) & $y$ (rudeness) \\
\hline
0.0 & no sound & undetectable \\
0.1 & whisper & indirect request: hint \\
0.2 & urgent whisper & preference\\
0.3 & subdued speech & query \\
0.4 & speaking voice & direct request: suggestion \\
0.5 & authoritative tone & obligation \\
0.6 & loud voice & command \\
0.7 & yell & generic foul words \\
0.8 & shout & targeted offense: eg. ethnic slur \\
0.9 & scream & \\
1.0 & shriek & threat of physical violence \\
\hline
\end{tabular}
\end{table}

Let us now consider the impact of differently parameterized $\alpha 10$ actions
on the self-perceived politeness of the client $v=${\tt\small~{\small \tt CSSM(Western, Politeness, Client, Crowd, Client)}}.

The perceived politeness can be either increased (for low values of $y$) or decreased (for high values of $y$). A louder voice can amplify the negative impact of rudeness, but it will not increase the politeness of mitigated speech. Furthermore, the impact $\Delta v$ will depend on the belief of the crowd with regards to whether the action involves reneging on an accepted transaction or whether it is the justifiable reproach addressed to a crooked seller, a value captured in the concrete belief $b=${\small \tt CB(S,Q-Agreed,Crowd,Client)}. The higher the belief that a commercial transaction has been agreed upon, the more negative impact the rudeness of the client will have on his perceived politeness. If the public perceives the seller as crooked, the rudeness of the client will be perceived as justifiable self-defence, and his perception will not suffer. On the other hand, the positive impact of polite behavior improves the metric regardless of the value of $b$ (one can be polite with a crooked seller).

Denoting with $v=${\small\tt CSSM(Western,Politeness,Client, Crowd,Client)}, we
have an AIF that can be modeled with the following logistic canonical form:

\noindent \begin{IEEEeqnarray}{lCl}
\Delta v & = & \big( L(y, -0.8, 0, 15) + L(1, 0.8, -100, 100)\big)\cdot \nonumber \\
         &   & ~~~\big( L(x, 50, 0, 0.08) + L(1, -25, -100, 100)\big) +\nonumber \\
         &   & L(y, -1, 0.95, 15)\cdot L(b, 1, 0.65, 8) \cdot \nonumber \\
         &   & ~~~\big( L(x, 50, 0, 0.08) + L(1, -25, 100, 100)\big) \nonumber
\end{IEEEeqnarray}

\begin{figure}\centering
  \includegraphics[width=0.45\textwidth]{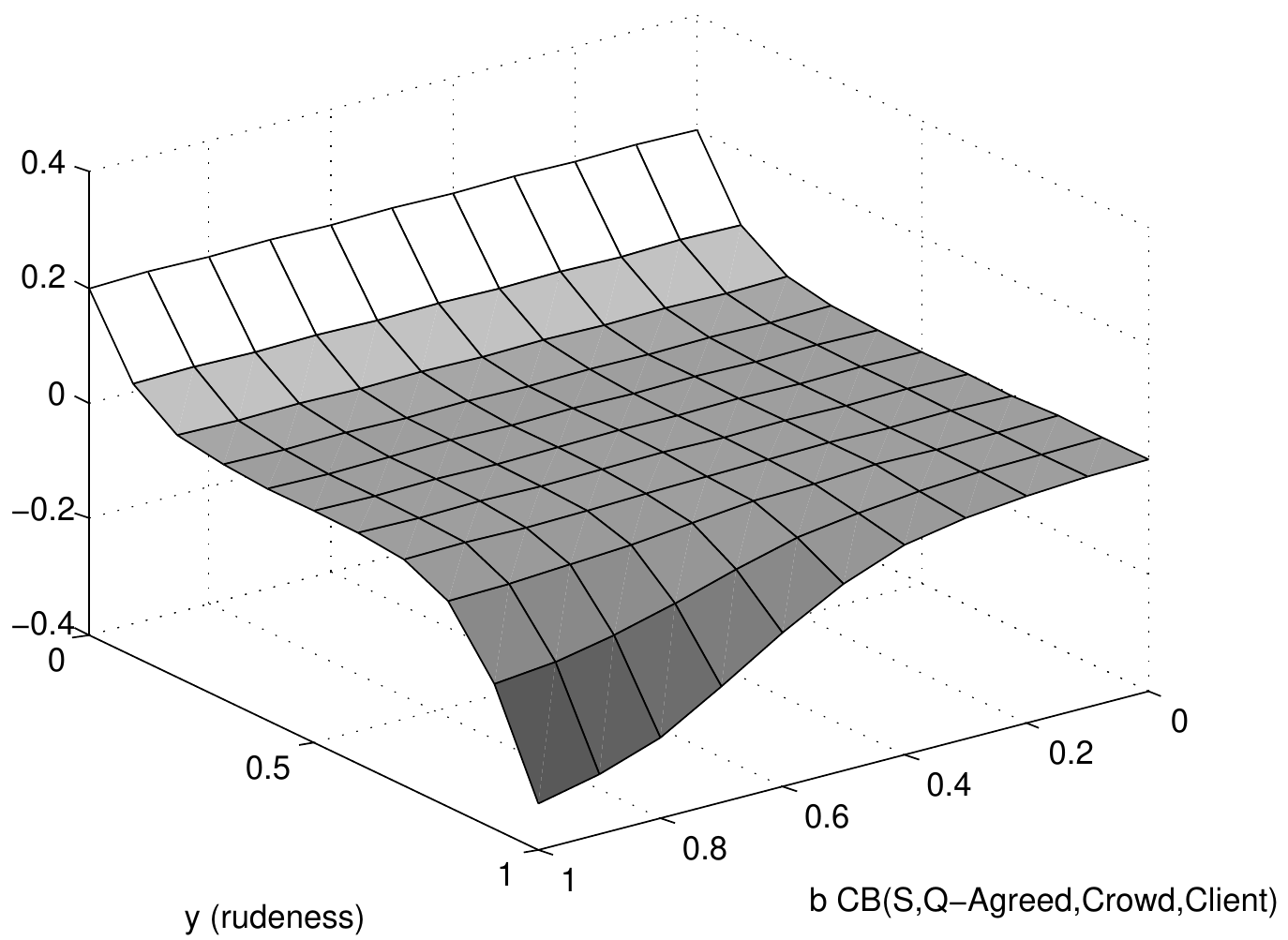}
  \caption{{\bf The impact of action {\small \tt CSSM(Western,Politeness, Client,Crowd,Client)} function of b and y for a fixed value of x=0.5}}
\label{fig:alpha10politeness}
\end{figure}

Figure~\ref{fig:alpha10politeness} shows that the evolution of $\Delta v$ function of the $b$ and $y$ values for a fixed value of $x=0.5$ indeed matches the informal description we provided above.

%
%
\subsection{CB AIFs in the Spanish Steps scenario}

In our model, actions affect the concrete beliefs through the application of the Dempster-Shafer conjunctive merge between a belief mass distribution representing the current belief and a belief mass distribution describing the weight of the evidence. To correctly track the evolution of the CBs we must associate a (possibly parameterized) belief mass distribution to every action.

\begin{table}
\begin{small}
\begin{center}
    \caption{{\bf Mass functions of evidence for {\tt CB(S, Q-Gift, Client, Client)}}}
    \label{tab:CBQGift}
\begin{tabular} {|p{3.2cm}|p{1.3cm}|p{1.8cm}|p{0.8cm}|}
\hline
{\bf Action} & {\tt\{T\}} & {\tt \{T,F\}} & {\tt \{F\}} \\
\hline
$\alpha 5$ (offers gift) & 0.3 & 0.7 & 0.0 \\
\hline
$\alpha 6$ (forces gift) & 0.3 & 0.7 & 0.0 \\
\hline
$\alpha 13(t)$ (waits) &
0.05 / sec & 0.95 / sec  & 0.0 \\
\hline
$\alpha 14$ (requests payment) & 0 & 0.0 & 1.0\\
\hline
$\alpha 16$ (concedes gift) & 1.0 & 0 & 0 \\
\hline
\end{tabular}
\end{center}
\end{small}
\end{table}

Table~\ref{tab:CBQGift} shows the belief mass distribution of various actions
affecting {\small \tt CB(S,Q-Gift,Client,Client)}. $\alpha 5$ is the action of
offering the flower as a gift, and it represents a weak evidence towards {\tt
\small Q-Gift} being true. $\alpha 6$ is the action of forcing the flower on the
client -- this can be interpreted either as an evidence for {\tt \small Q-Gift},
but also towards its opposite. Both mass distributions keep significant
uncertainty. Depending on the belief the agent started from, after these actions
the client might still be mostly inclined to believe the flower not to be a
gift. Every second that the seller leaves the client with the flower without
asking for money (action $\alpha 13$) provides more evidence towards the flower
being a gift. Action $\alpha 14$ requesting payment will
immediately clarify that the flower is not a gift, and will reduce the uncertainty to zero. In contrast, action $\alpha 16$, conceding the gift, will set the CB to 1.0, also reducing the uncertainty to zero. This action, however, is only a fictional one, which might be expected by an uninformed client, but will never be performed by the seller.

%
%
\subsection{Case study 1: Successful sell}

In the following, we model two real-world scenario instances witnessed on July 6, 2012 at the Spanish Steps, Rome, Italy.

In the first observed scenario the seller was successful in selling the flower to a romantic couple. The seller offered the bouquet to the man ($\alpha 1$), but was declined ($\alpha 4$). Then, the seller offered a flower to the woman ($\alpha 5$), and she accepted it. After a waiting time of 15 seconds some distance away ($\alpha 13(15)$), the seller returned and requested payment from the man ($\alpha 14$). The client attempted to return the flower, with low voice and suggestion type mitigation level ($\alpha 10(0.2,0.4)$). The seller declined to take back the flower ($\alpha 11$). At this point, the man accepted to pay ($\alpha 2$) and paid for the flower ($\alpha 3$).

\begin{center}
\noindent TS $\xrightarrow{\alpha 1}$ S1
$\xrightarrow{\alpha 4}$ S3
$\xrightarrow{\alpha 5}$ S4
$\xrightarrow{\alpha 7}$ S7
$\xrightarrow{\alpha 13(15)}$ S7
$\xrightarrow{\alpha 14}$ S8
$\xrightarrow{\alpha 10(0.2,0.4)}$ S9
$\xrightarrow{\alpha 11}$ S8
$\xrightarrow{\alpha 2}$ S10
$\xrightarrow{\alpha 3}$ TP2
\end{center}

\begin{figure*}
\begin{center}
\begin{tabular}{c}
\includegraphics[width=0.4\textwidth]{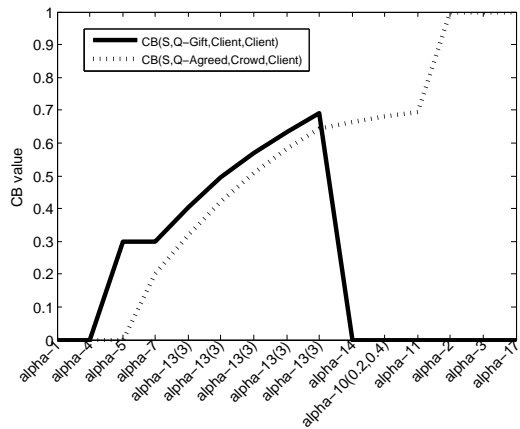} \\
(a) concrete beliefs \\
\end{tabular}
\begin{tabular}{cc}
\includegraphics[width=0.4\textwidth]{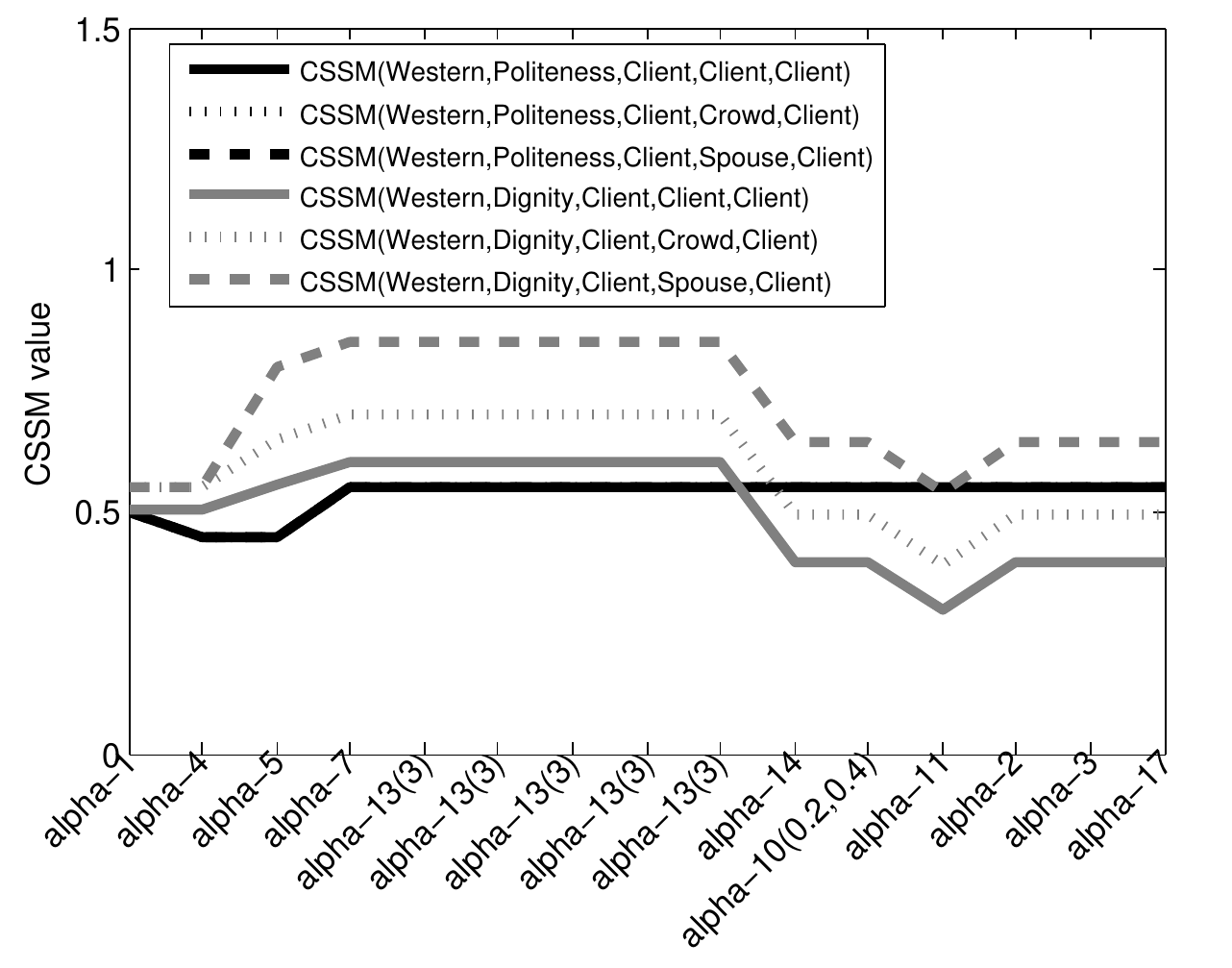} &
\includegraphics[width=0.4\textwidth]{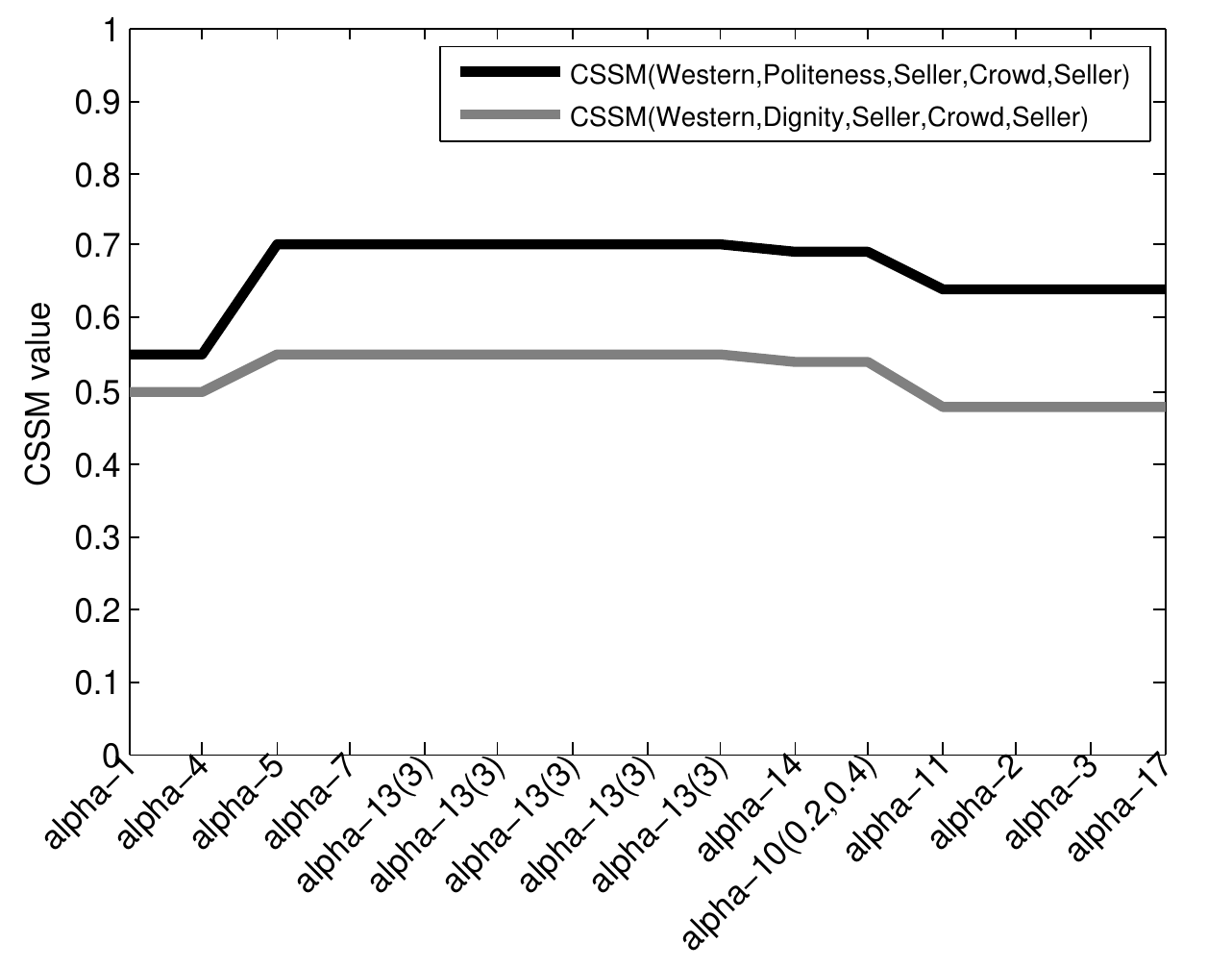} \\
 (b) client politeness and dignity & (c) seller politeness and dignity \\
\includegraphics[width=0.4\textwidth]{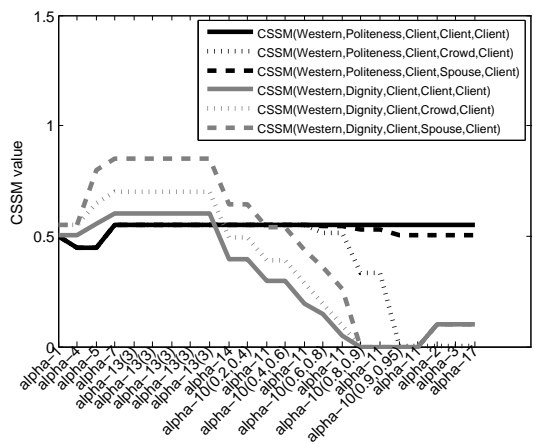} &
\includegraphics[width=0.4\textwidth]{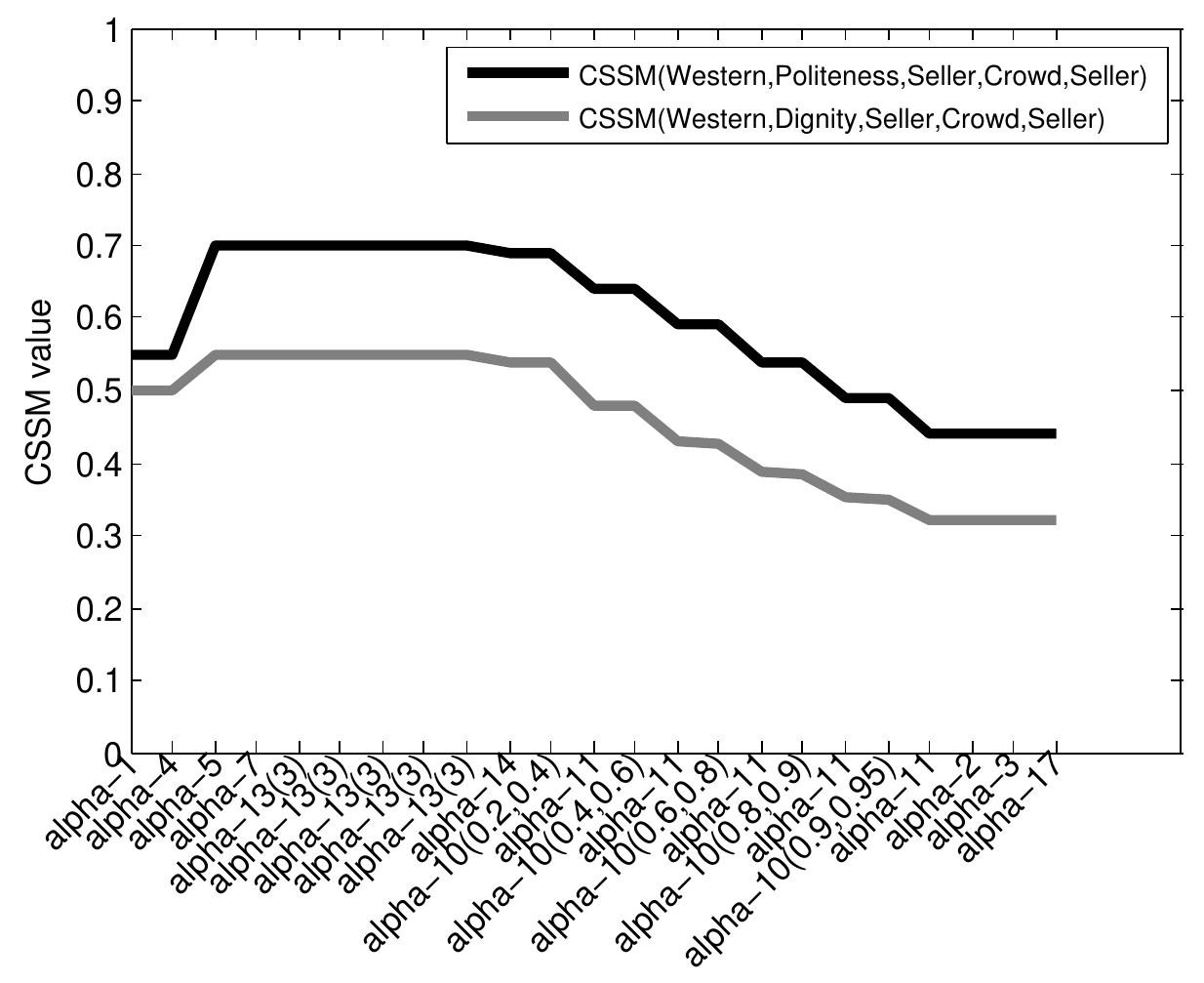} \\
(d) client: what would have happened? & (e) seller: what would have happened?  \\
\end{tabular}
\end{center}
\caption{{\bf Scenario 1 - Successful sell} : (a) the evolution of the CBs {\tt Q-Gift} and {\tt Q-Agreed}, (b) the politeness and dignity CSSMs of the client, (c) the politeness and dignity CSSMs of the seller (d) the politeness and dignity CSSM of the client in a fictional ``what would have happened?'' scenario (e) the politeness and dignity CSSM of the seller in a fictional ``what would have happened?'' scenario }
\label{fig:Scenario1}
\end{figure*}

What requires explanation in this scenario is the fact that the client gives in relatively easily, despite the fact that he does not want the flower (as he tries to return it) and he knows that he is being cheated. Figure~\ref{fig:Scenario1} shows the results of tracking this scenario using our model. For all the graphs, the X axis lists the actions and their parametrization.

Figure~\ref{fig:Scenario1}-a shows the evolution of the concrete beliefs. The {\small \tt CB(S,Q-Gift,Client,Client)} starts with a zero value, then it raises to about 0.32 after $\alpha 5$ (the offering of a single flower). This appears to be sufficient for the client to accept the flower as a gift. Albeit this value appears to be low, note that this is the Dempster-Shafer belief value which does not imply that the client has a 0.68 belief in the fact that the flower is {\em not} a gift - the majority of the remainder of the belief mass is concentrated in the uncertainty domain $\{T,F\}$. The belief that the flower is a gift will actually climb during the waiting time of action $\alpha 13$ which means that if the client did not give back the flower initially, it will be unlikely that he will give it back during this wait\footnote{This statement assumes that no other event changes the client's belief throughout the wait. We have witnessed scenarios where the client holding the flower had seen another client being asked for money, and rushed to return the flower himself, illustrating how actions in one scenario can change CBs in another. This situation can be modeled by our framework but it is beyond the scope of the examples considered in this section.}. The belief that the flower is a gift plummets to zero once the client is asked for money.

Let use now see the evolution of the client's estimate of the crowd's belief that a transaction had been agreed upon {\small \tt CB(S,Q-Agreed,Crowd,Client)}. {\small \tt Q-Agreed} only tracks the existence of an agreement about a transaction -- the actual nature of the transaction changes: up to $\alpha 14$ the client believes that the transaction had been gift giving, after $\alpha 14$, it is clear that the transaction is a commercial one.

From an initial value of 0, this CB jumps to a value of about 0.25 once the client accepts the flower, and gradually increases as long as the client holds the flower. Whether this is a good estimate of the crowd's belief has no relevance to the scenario as long as the crowd is passive\footnote{It is not impossible for the crowd to become an active participant in a scenario - people might intervene verbally or call the police.}. In fact, it is quite likely that the majority of the crowd members did not notice or follow the transaction.

What is relevant from an explanation and prediction point of view is the fact that at the moment when the client is asked for money and makes his attempt to return the flower, this CB has a relatively high value (about 0.68).

Figures~\ref{fig:Scenario1}-b and \ref{fig:Scenario1}-c track the evolution of the dignity and politeness metrics of the client and the seller. Overall, this particular scenario was a very polite interaction, thus we see only moderate changes in the politeness values. The dignity of the client sees somewhat more variation - it initially increases (when the client believes that his spouse is being honored with a gift) and then decreases - when he realizes that he is being cheated. Overall, the client finishes the scenario with quite high dignity and politeness CSSMs. On the other hand, he was obviously cheated and suffered a financial loss.

From an explanatory and predictive perspective, the question is: why did the client accept to pay for the flower? Could we have predicted this outcome? To answer this we can now create a ``what would have happened'' scenario, where we follow the observed scenario up to a point, and then change it to see what would happen if the client makes a different decision. Figures~\ref{fig:Scenario1}-d and ~\ref{fig:Scenario1}-e shows the client's and seller's dignity and politeness in a scenario where, instead of deciding to pay after the first return attempt, the client escalates his return attempts using louder and louder voice and increasingly rude language and gestures. What we see is that this scenario quickly leads to a catastrophic decay of both the dignity and the crowd-perspective politeness of the client while the public politeness and dignity of the seller had been barely impacted. This asymmetry is due to the fact that the client performs these acts in public in front of a crowd appearing to believe that he is reneging an accepted transaction.

The {\em explanation} for the client paying is that he could continue his return efforts only at a very high cost for his public politeness and dignity while the seller can afford to decline the return of the flower with minimal impact to his public perception.

Similarly, after action $\alpha 13$ our model allows us to {\em predict} that the selling action will be successful, as long as we are only considering the calculations of CSSMs and assume that the humans attempt to maximize them. While many people follow social norms as represented by the CSSMs, some of
them are not: we witnessed cases where the client has thrown the flower to the
ground and walked away in anger.

%
%
\subsection*{Case study 2: Unsuccessful sell}

\begin{figure*}
\begin{center}
\begin{tabular}{c}
    \includegraphics[width=0.40\textwidth]{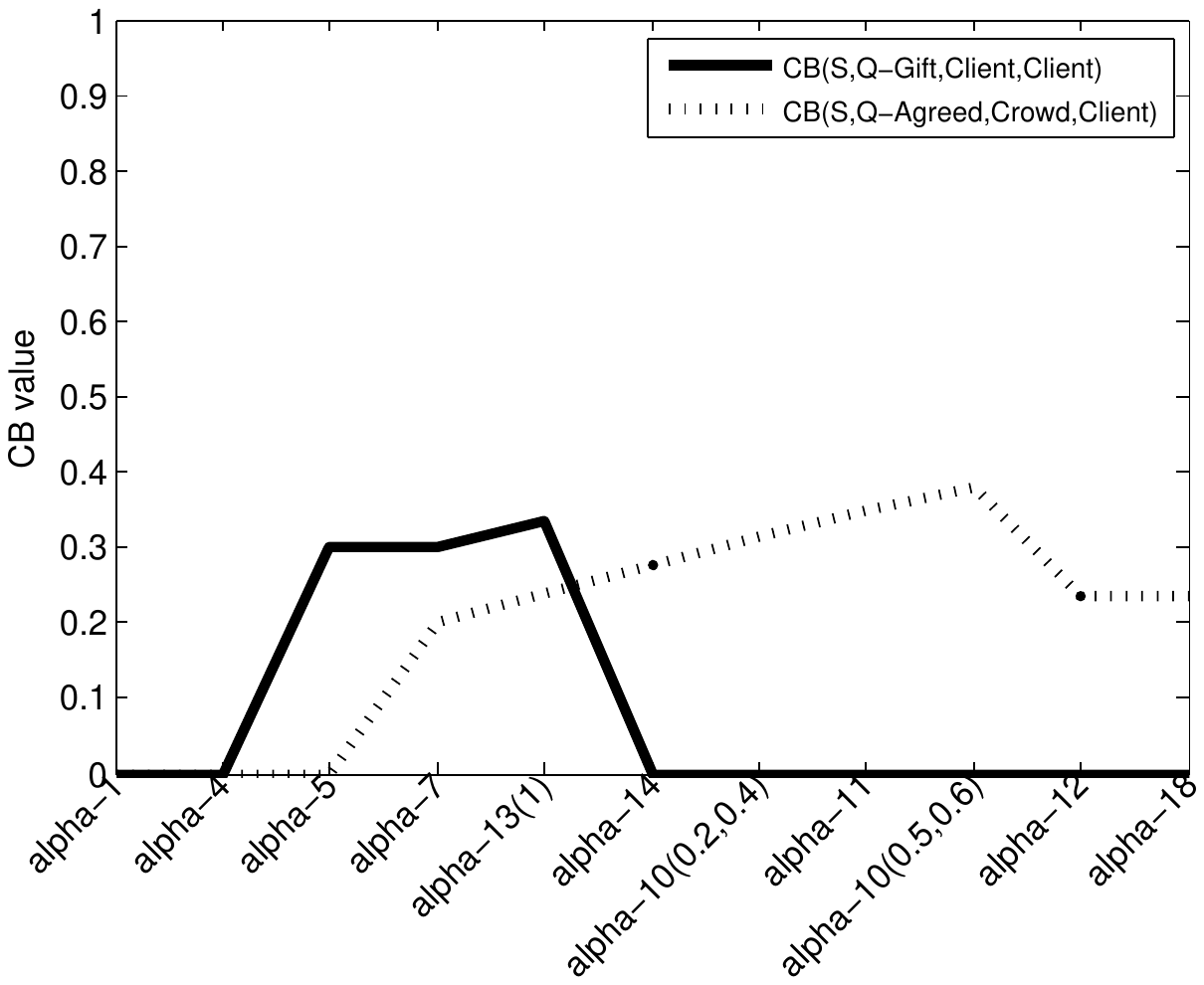} \\
    (a) concrete beliefs
\end{tabular}
\begin{tabular}{cc}
\includegraphics[width=0.4\textwidth]{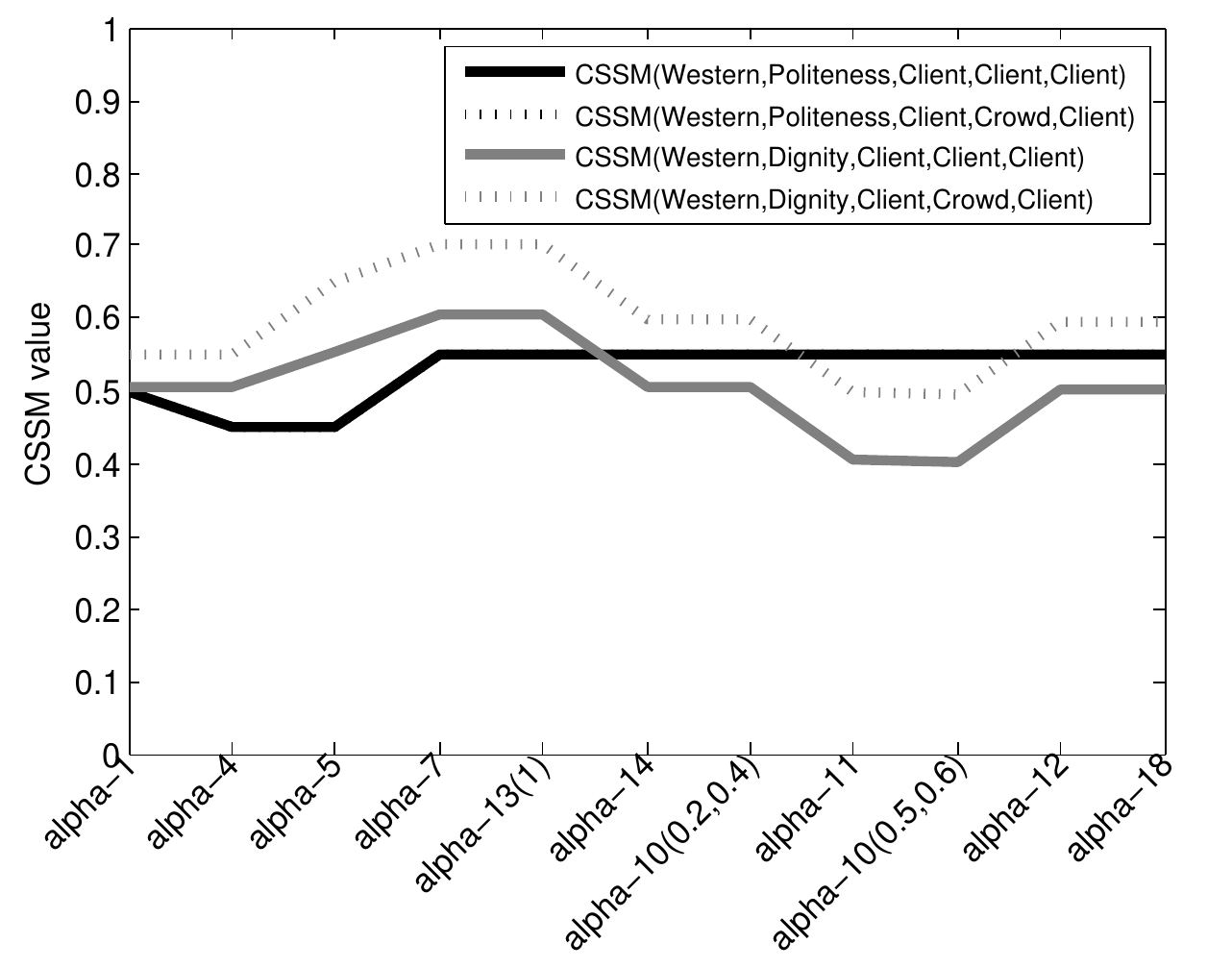} &
\includegraphics[width=0.4\textwidth]{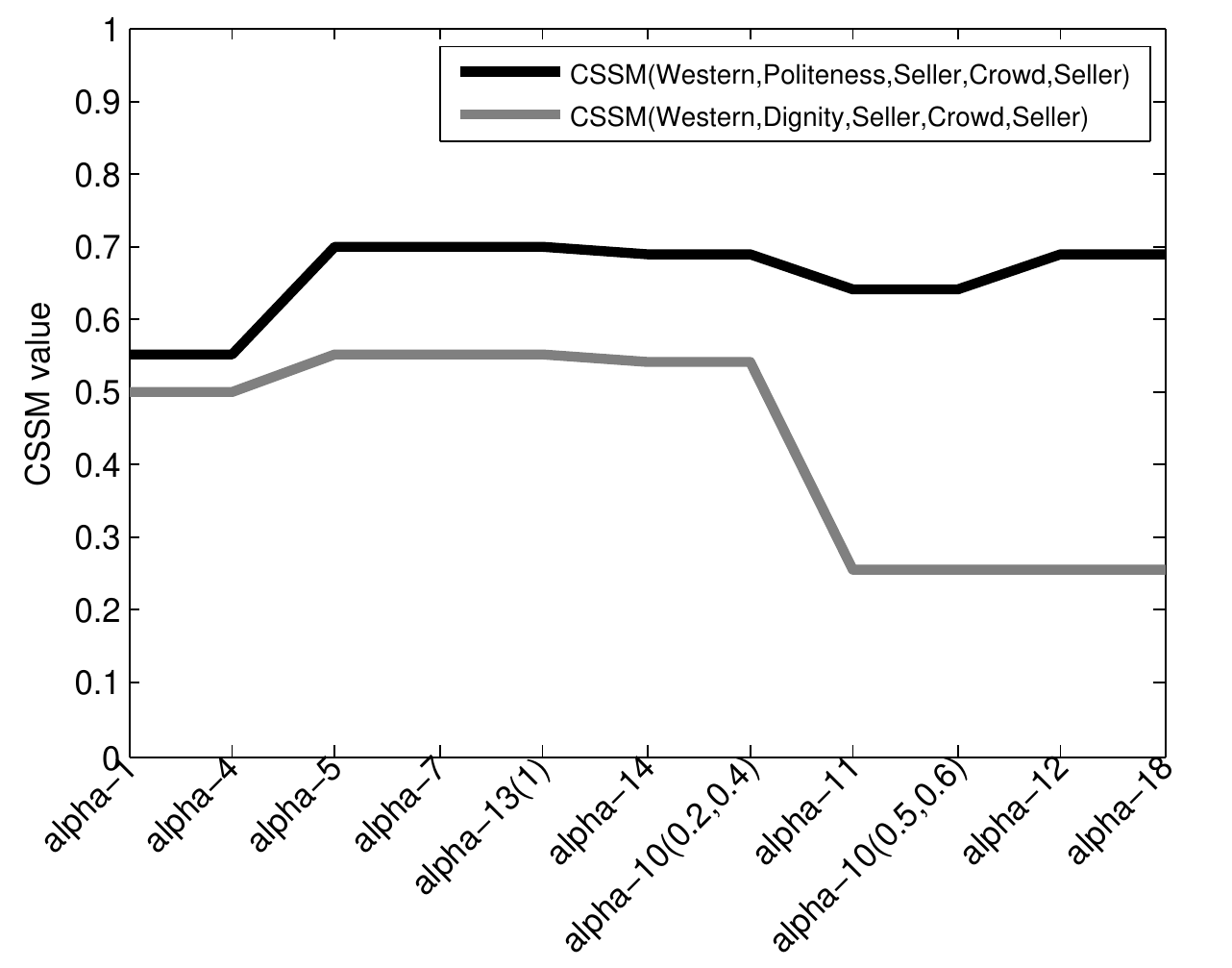} \\
 (b) client politeness and dignity & (c) seller politeness and dignity \\
\includegraphics[width=0.40\textwidth]{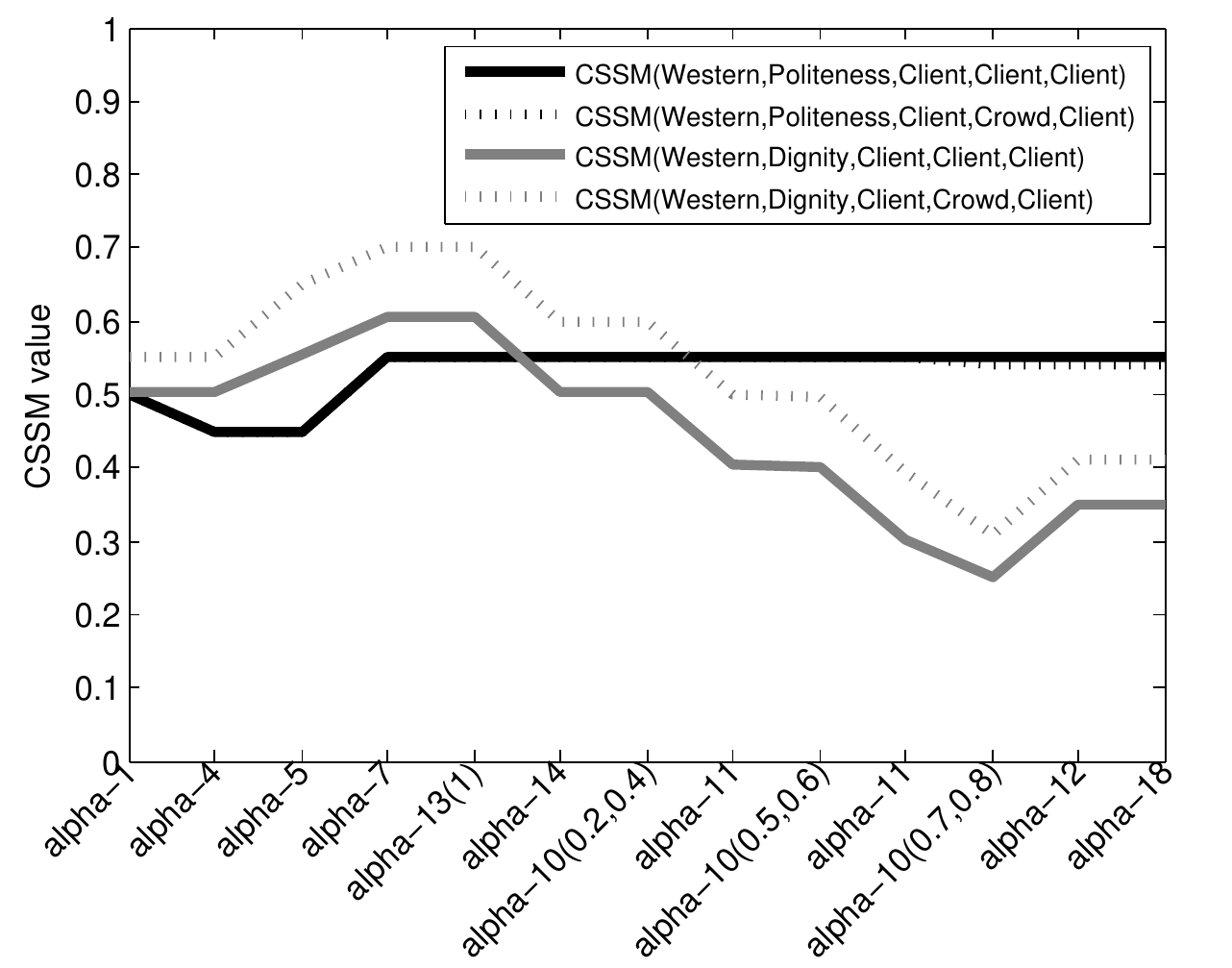} &
\includegraphics[width=0.40\textwidth]{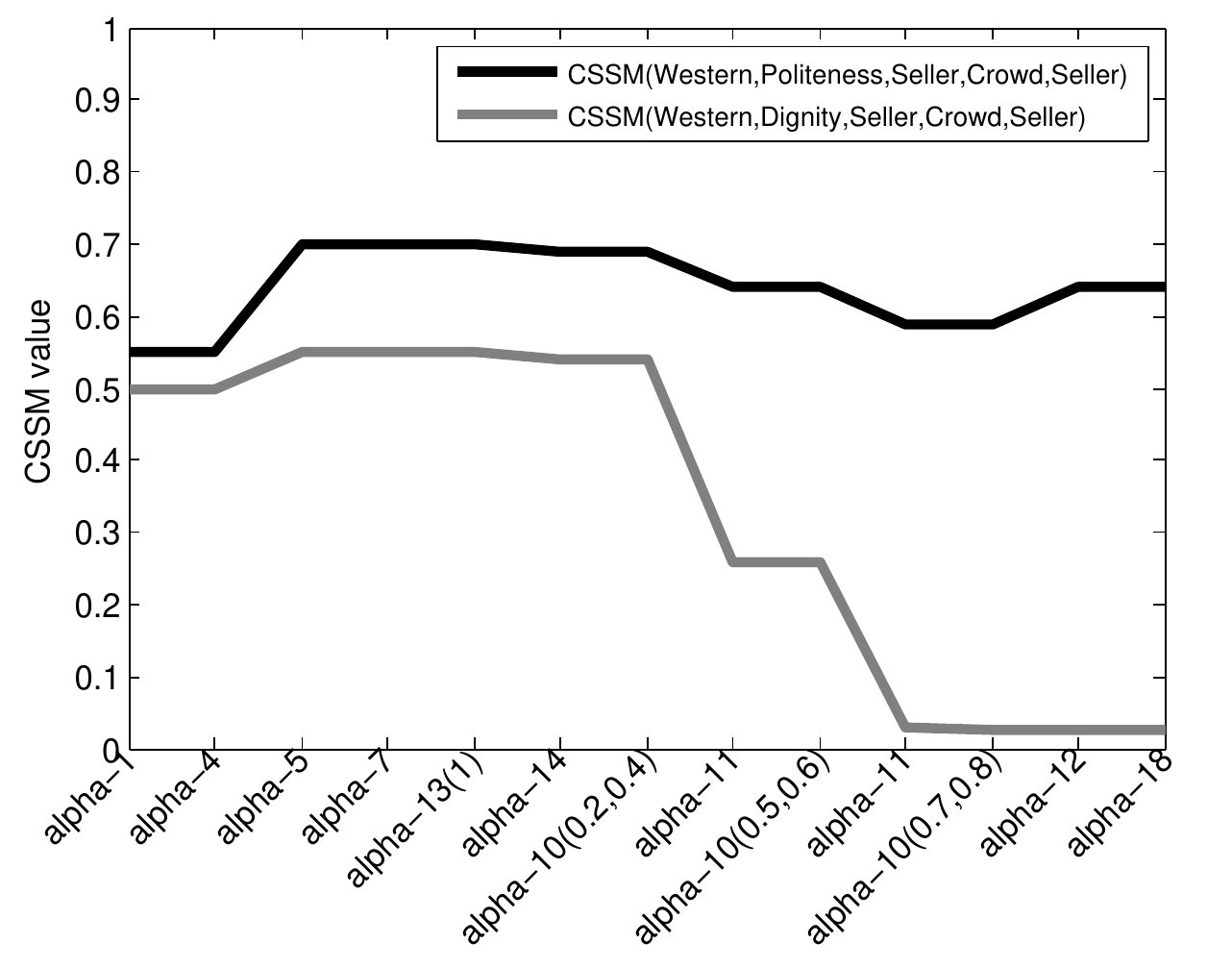} \\
(d) client: what would have happened? & (e) seller: what would have happened?\\
\end{tabular}
\end{center}
\caption{{\bf Scenario 2 - Unsuccessful sell}: (a) the evolution of the CBs {\tt Q-Gift} and {\tt Q-Agreed}, (b) the politeness and dignity CSSMs of the client, (c) the politeness and dignity CSSMs of the seller (d) the politeness and dignity CSSM of the seller in a fictional ``what would have happened?'' scenario (e) the politeness and dignity CSSM of the seller in a fictional ``what would have happened?'' scenario}
\label{fig:Scenario2}
\end{figure*}

The second scenario shows an instance where the seller was unsuccessful in selling the flower to a single woman. The start of the scenario was similar to the previous case. However, as the woman moved to leave the area, the seller asked her for money only one second after the flower was accepted. The woman had attempted a return in polite terms and low voice $\alpha 10$(0.2,0.4). After the return was declined, the woman in firm terms but without using expletives ordered the vendor to take back the flower ($\alpha 10$(0.5,0.6)). At this point the vendor accepted the return ($\alpha 12$).

\begin{center}
\noindent TS $\xrightarrow{\alpha 1}$ S1
$\xrightarrow{\alpha 4}$ S3
$\xrightarrow{\alpha 5}$ S4
$\xrightarrow{\alpha 7}$ S7
$\xrightarrow{\alpha 13(1)}$ S7
$\xrightarrow{\alpha 14}$ S8
$\xrightarrow{\alpha 10(0.2,0.4)}$ S9
$\xrightarrow{\alpha 11}$ S8
$\xrightarrow{\alpha 10(0.5,0.6)}$ S9
$\xrightarrow{\alpha 12}$ TN2
\end{center}

Figure~\ref{fig:Scenario2}-a, b and c shows the evolution of the CBs, the client's politeness and dignity and the seller's politeness and dignity respectively. To avoid unnecessary repetitions we will concentrate on the differences from the successful sell scenario. The first observation is that the client being a single woman, the spouse-perspective values are not present in the client's evaluation.

In the CBs the main difference is that as the seller was in a rush to ask for money, the asking for money $\alpha 14$ happens at a much lower value of {\small \tt CB(S,Q-Agreed,Crowd, Client)}.

With regards to the CSSMs, both participants end up with a lower dignity. The
politeness, however, is relatively unaffected: the client does not use very rude
words and gestures (and is protected by the fact that the {\small \tt Q-Agreed}
CB is relatively low). The seller looses some politeness by his first refusal to take back the flower, but recovers in politeness when it accepts the return. He looses relatively large measures of dignity by his refusal. The reason for this is that his estimate of the public belief in an agreement is the same as the client's: {\small \tt CB(S,Q-Agreed,Crowd,Seller) = CB(S,Q-Agreed,Crowd,Client)}. This means that in his estimate, the crowd is more likely to see this as a forced transaction by a crooked seller, which would make his refusal have a larger impact. Overall, the seller finishes the scenario with acceptable values of dignity and politeness. On the other hand, he did not make a profit and wasted time.

From an explanatory and predictive perspective the question we must ask is why
the seller gives up in this particular scenario, and whether it was possible to
predict this outcome. Again, we will create a ``what would have happened'' scenario, where we assume that the seller, instead of giving in, would have repeatedly declined the return (action $\alpha 11$) in the face of more and more insistent return efforts from the client. The client's and seller's CSSMs for this hypothetical scenario are shown in Figures~\ref{fig:Scenario2}-d and \ref{fig:Scenario2}-e.

What we find in this case is that the seller would suffer a socially
unacceptable decline in public dignity. On the other hand, the client had only a moderate decrease in the dignity and politeness during this escalation phase. This explains why the seller decided to give up the transaction without insisting further.

From a prediction point of view, after $\alpha 13$ and $\alpha 14$, that is after the seller asks for money prematurely, we can predict that the sell will likely fail, because the client can push the seller into deep public dignity loss while suffering relatively minor damage to her own politeness and dignity. Again, this prediction is probabilistic and depends on the willingness of the client to start the escalation of the return effort - if the client gives up after the first try, the scam can still succeed.

%
%
\section{Conclusions}
\label{sec:Conclusions}

In this paper we described an approach to create a computational model of social norms based on identifying values that a certain culture finds desirable. The model quantifies these values in the form of Culture-Sanctioned Social Metrics (CSSMs) and sees social norms as the requirement to maximize these metrics from the perspective of the self, peers and public. Our model uses action-impact functions (AIFs) to represent the way actions affect the CSSMs, this being sometimes also dependent on the concrete beliefs (CSs) of the actors about specific questions on the social scenario. Beyond the generic model, we also proposed some specific representation choices suitable for the handcrafting of the AIFs: the sum of products of logistic functions form for CSSM AIFs and the Dempster-Shafer model for CB AIFs. We validated the model by using it to represent a complex deception scenario and showing that it can yield non-trivial insights such as the explanation of apparently irrational human behavior.

We are working in several directions to further improve the model. We are continuously extending the collection of modeled cultures and scenarios.
To facilitate this, we are investigating the possibility to replace the handcrafting of the CSSM and CB AIFs from observations and interviews with the automated acquisition of them from a small number of scenarios. This requires the development of new AIF models possibly based on deep neural networks.

\section*{Acknowledgments}

Acknowledgements: The research reported in this document/presentation was performed in connection with Contract Number W911NF-10-2-0016 with the U.S. Army Research Laboratory. The views and conclusions contained in this document/presentation are those of the authors and should not be interpreted as presenting the official policies or position, either expressed or implied, of the U.S. Army Research Laboratory, or the U.S. Government unless so designated by other authorized documents. Citation of manufacturer's or trade names does not constitute an official endorsement or approval of the use thereof. The U.S. Government is authorized to reproduce and distribute reprints for Government purposes notwithstanding any copyright notation heron.

\bibliographystyle{abbrv}
\bibliography{References}

\end{document}